\documentclass[twocolumn]{aastex631}
\usepackage{amsmath}
\usepackage{graphicx}

\begin{document}

\title{Creating the Radius Gap without Mass Loss}

\correspondingauthor{Eve J.~Lee}
\email{evelee@physics.mcgill.ca}

\author[0000-0002-1228-9820]{Eve J.~Lee}
\affil{Department of Physics and McGill Space Institute, McGill University, Montr\'eal, Qu\'ebec, H3A 2T8, Canada}
\affil{Institute for Research on Exoplanets (iREx), Montr\'eal, Qu\'ebec, Canada}

\author[0000-0003-0525-1805]{Amalia Karalis}
\affil{Department of Physics and McGill Space Institute, McGill University, Montr\'eal, Qu\'ebec, H3A 2T8, Canada}
\affil{Institute for Research on Exoplanets (iREx), Montr\'eal, Qu\'ebec, Canada}

\author[0000-0002-5113-8558]{Daniel P.~Thorngren}
\affil{Institute for Research on Exoplanets (iREx), Montr\'eal, Qu\'ebec, Canada}
\affil{Department of Physics, University of Montr\'eal, Montr\'eal, Qu\'ebec, Canada}

\begin{abstract}
The observed exoplanet population features a gap in the radius distribution that separates the smaller super-Earths ($\lesssim$1.7 Earth radii) from the larger sub-Neptunes ($\sim$1.7--4 Earth radii). While mass loss theories can explain many of the observed features of this radius valley, it is difficult to reconcile them with potentially rising population of terrestrials beyond orbital periods of $\sim$30 days. We investigate the ability of gas accretion during the gas-poor phase of disk evolution to reproduce both the location of the observed radius gap and the existence of long-period terrestrial planets. Updating the analytic scaling relations of gas accretion rate accounting for the shrinking of the bound radius by hydrodynamic effects and deriving a more realistic disk temperature profile, we find that the late-stage gas accretion alone is able to carve out the observed radius gap, with slopes $R_{\rm gap} \propto P^{-0.096}$ and $R_{\rm gap} \propto M_\star^{0.15}$ for top-heavy; and $R_{\rm gap} \propto P^{-0.089}$ and $R_{\rm gap} \propto M_\star^{0.22}$ for bottom-heavy core mass distributions, in good agreement with observations. The general morphology of the primordial radius gap is stable against a range of disk gas density and disk accretion rate with the latter affecting mostly the population of large planets ($\gtrsim$3--4$R_\oplus$). The peaks and valleys in the radius distribution were likely set in place primordially while post-formation mass loss further tunes the exoplanetary population. We provide potential observational tests that may be possible with TESS, PLATO and Roman Space Telescope.
\end{abstract}

\section{Introduction}
\label{sec:intro}

Out of all the detected exoplanets thus far, super-Earths and sub-Neptunes comprise the large majority, with approximately $\sim$30--50\% of Sun-like stars harbouring at least one of these small planets within orbital periods of $\sim$300 days \citep[e.g.,][]{fressin13,petigura13,burke15,Zhu18}. From measurements of radii and masses alone, interior modeling shows degeneracy between different compositions of the envelope and the planetary cores, which made unclear where to draw the line between the scaled-up version of rocky Earth and the scaled-down version of gas-enveloped Neptune \citep[e.g.,][]{rogers10-general}. \citet{Rogers15} proposed that the distinction should be made at around $\sim$1.6$R_\oplus$ beyond which some level of H/He-dominated gaseous layer is required to explain both the masses and radii of close-in {\it Kepler} planets. With a larger dataset and more precise measurements of stellar parameters, we now have clear observational evidence of the distinction between super-Earths and sub-Neptunes which are evinced by a clear gap in the radius and radius-period distributions \citep[e.g.,][]{Fulton17,vanEylen18, Martinez19,Petigura20,David21}.

This radius valley is considered to be a signature of envelope mass loss whether by photoevaporation \citep{Owen13,Owen17} or by core-powered envelope mass loss \citep{Ginzburg18,Gupta19}. In these theories, all super-Earths begin as gas-enveloped sub-Neptunes, and those that are most susceptible to mass loss (i.e., those with low gravity and close proximity to the star) shed their outer envelope, transforming into bare rocks. Both theories enjoy success in explaining many of the observed features of the radius gap including its location and its overall shape in both one-dimensional radius and two-dimensional radius-period distributions, as well as the shift of the gap with respect to host stellar mass. Where these two theories differ is the range of wavelengths of the stellar flux that matters. For photoevaporation, only the XUV spectrum is relevant whereas for core-powered envelope mass loss, it is the bolometric flux that is key. These differences will materialize in the dependence of the radius gap with respect to stellar mass at a fixed bolometric incident flux; unfortunately, current dataset is not big enough to test this \citep{Loyd20,Rogers21_comb}. 

Another potential way to distinguish between the two mass loss models is the system age. It is generally expected that photoevaporative mass loss carves out the gap in the first $\sim$100 Myrs when the stars remain active whereas core-powered envelope mass loss is expected to carve out the gap over a longer $\sim$Gyr timescale. There are observational evidence that the radius distribution shifts at a statistically significant level across Gyr timescales \citep[e.g.,][]{Berger20,David21}. However, EUV luminosity is expected to decline more slowly compared to X-ray so that photoevaporative mass loss may operate over $\sim$Gyrs as well \citep[e.g.,][]{King21}. 

\begin{figure}
    \centering
    \includegraphics[width=0.5\textwidth]{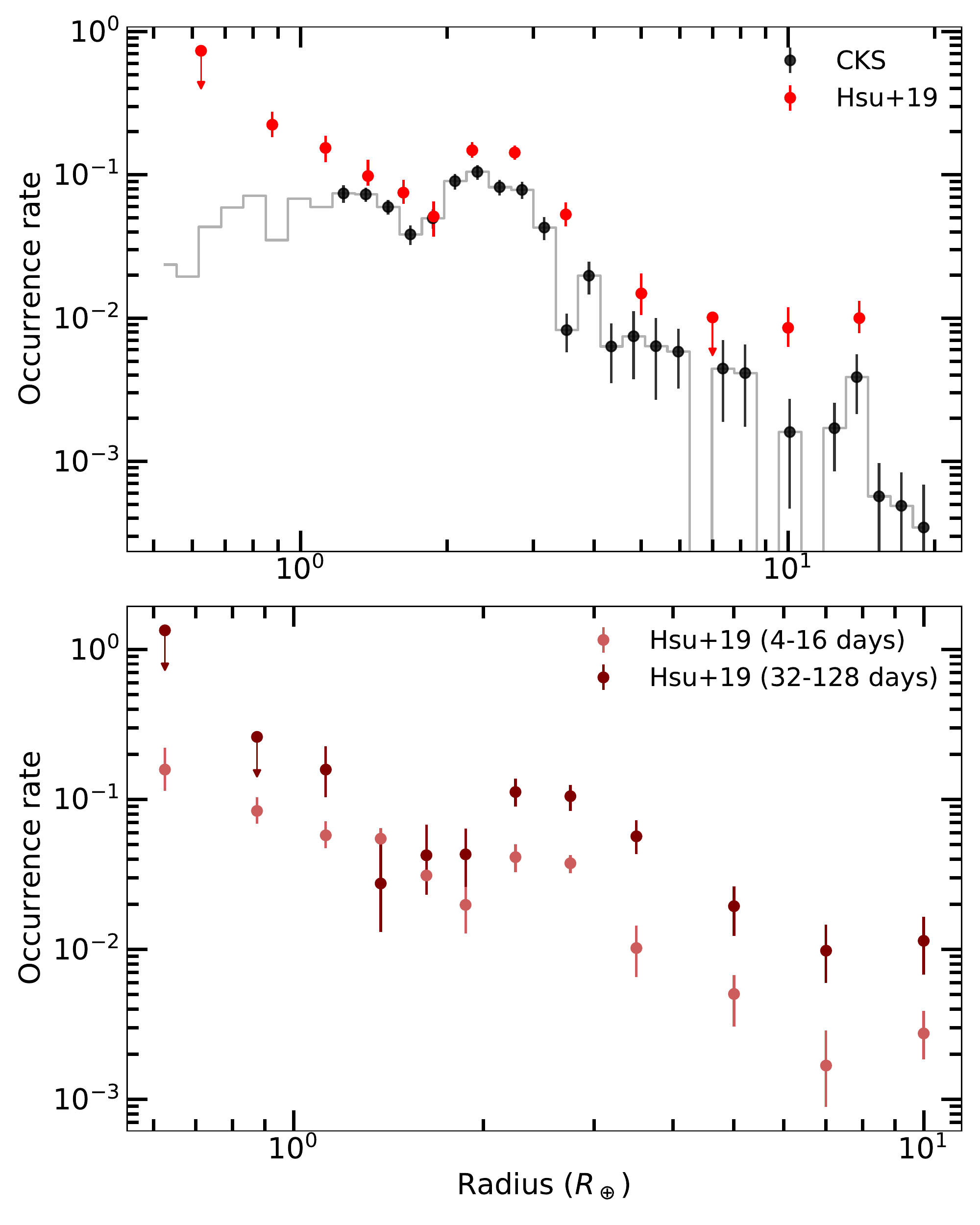}
    \caption{Top: Comparison of occurrence rates studies performed by California Kepler Survey \citep[][$<$100 days; their Figure 5]{Fulton18} and by \citet[][2--64 days; their Table 2]{Hsu19}. The downward arrows denote upper limits. While the overall shape and the approximate location of the radius gap are similar between the two studies, \citet{Hsu19} find overall broader radius distribution both towards larger radii ($>3R_\oplus$) and smaller radii with a hint of sub-Earths ($<R_\oplus$) becoming a dominant population. Bottom: Radius distribution at different orbital periods from \citet{Hsu19}. We verify their finding that there is no evidence of decrease in small planet ($\lesssim$1.4$R_\oplus$) population at long orbital periods.}
    \label{fig:obs_rdistrb}
\end{figure}

Alternatively, the origin of the radius gap may trace to the different composition of cores. \citet{Zeng19} have argued that the radius gap manifests in the mass-radius space separating $\sim$1--10$M_\oplus$ + $\lesssim$2$R_\oplus$ objects from $\sim$3--20$M_\oplus$+$\sim$2--3$R_\oplus$ objects whereby the former represents silicate+Fe rocky terrestrials and the latter represents icy planets \citep[see also][]{Venturini20}. Interpreting the radius gap as a compositional dichotomy is hard to reconcile however with the observed trend between the location of the radius gap with orbital period/stellar insolation and stellar mass as reported by \citet{Berger20}, and it may be that Earth-like cores with a large variety of H-atmosphere content suffice to explain the two distinct groups of planets that can be observed in the density space \citep[see, e.g.,][their Figure 21]{David21}. This does not however rule out the possibility that at least some of the super-Earth/sub-Neptune cores are icy rather than pure rocks, bearing in mind that the number of planets with measured masses is far smaller than the number of planets with radii measurement and so at present, it is difficult to make one-to-one correlation between the gap seen in radius-period space and the gap seen in mass-radius space.

Irrespective of the composition of cores, \citet{Lee21} proposed that the radius gap may be carved out during late-time gas accretion, well before mass loss. For every core mass at a given orbital period, there exists a maximum possible mass the core can accrete, set by the isothermal maximally cooled limit \citep{Lee15}. Cores lighter than $\sim$1--2$M_\oplus$ (with the actual mass depending on orbital period) are so tiny that their isothermal limit attains envelope mass fractions of $\lesssim 10^{-4}$. With such tiny amount of gas, these planets' radii are dominated by the size of their cores. The sharp drop of the expected envelope mass fraction below certain core masses manifests as a primordial gap in the radius and radius-period distribution with their location and shape consistent with what is currently observed. 

One signature of this primordial radius gap is the population of small planets at orbital periods beyond $\sim$30 days. At these orbital periods, both photoevaporation and core-powered envelope mass loss loses its potency and so these theories would expect very few of these planets at longer orbital periods.\footnote{Such prospect implies that current mass loss theories would predict solar system terrestrials to be rare or be a secondary population \citep{Owen17}.} There are already hints of the existence and dominance of super-Earths and sub-Earths beyond $\sim$30 days \citep[][see also our Figure \ref{fig:obs_rdistrb}]{Hsu19}.

\begin{figure}
    \centering
    \includegraphics[width=0.5\textwidth]{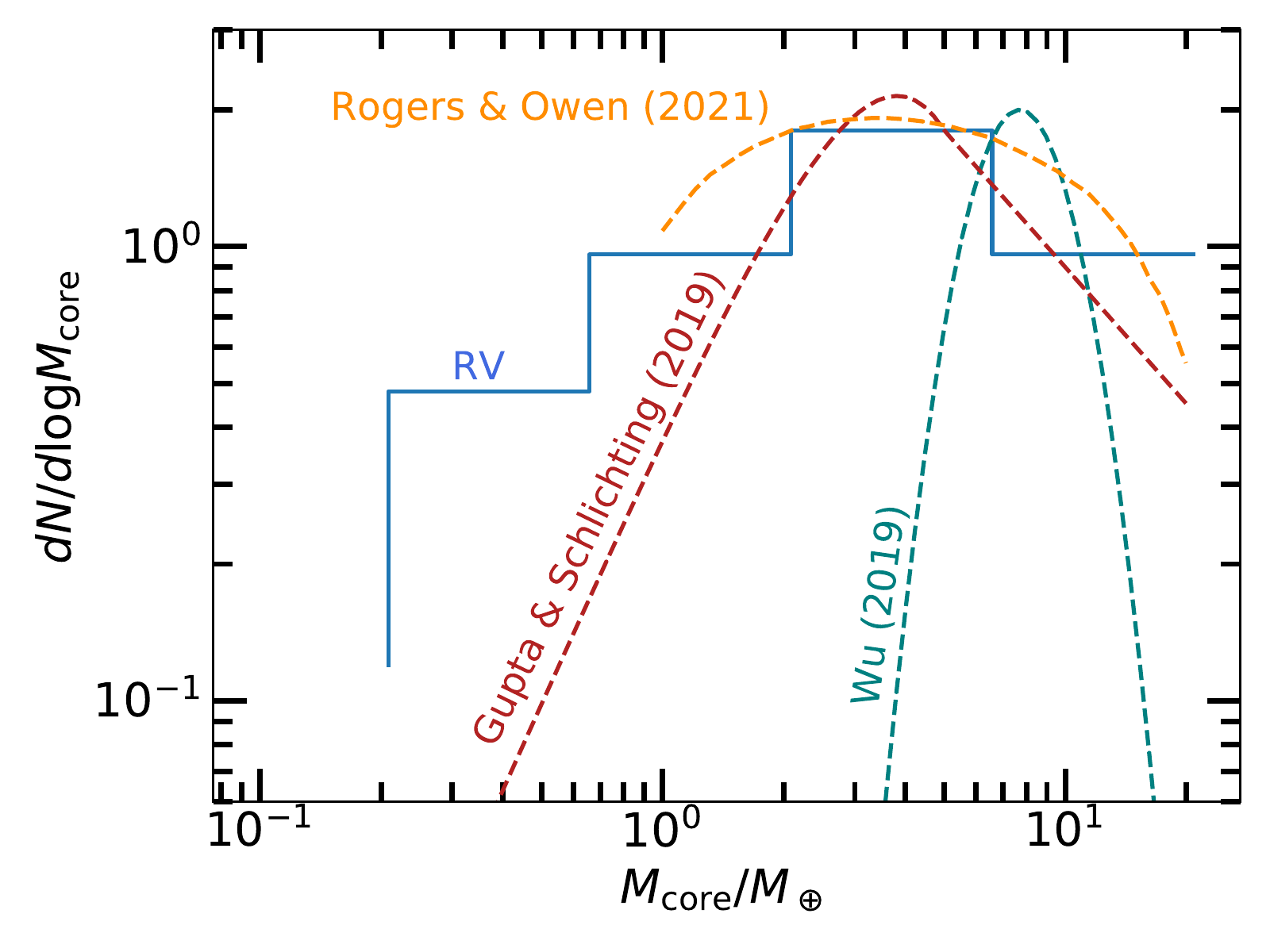}
    \caption{Best-fit core mass distributions of mass-loss models compared with the radial velocity follow-up of {\it Kepler} sub-Neptunes (RV, blue histogram) from \citet{Marcy14}. The RV distribution is not corrected for any observational biases. \citet{Wu19} assumes a lognormal core mass distribution and fits the photoevaporative mass loss model to the California Kepler Survey (CKS) radius distribution from \citet{Fulton18}. They find a distribution that is too peaked and shifted towards too high a core mass. \citet{Gupta19} assume a Rayleigh distribution for $M_{\rm core} < 5M_\oplus$ and $dN/dlogM_{\rm core} \propto M_{\rm core}^{-1}$ for larger masses and fit the core-powered envelope mass loss model to the CKS data. Their lower mass tail falls off too sharply. \citet{Rogers21_pevap} consider a non-parametric form of $dN/dlogM_{\rm core}$ fitting the photoevaporative loss model to the CKS data, finding a much broader distribution. They still expect a fall-off at $M_{\rm core} < M_\oplus$.}
    \label{fig:dNdMc}
\end{figure}

Another advantage of the primordial radius gap is that it can accommodate broad core mass distributions. As shown in Figure \ref{fig:dNdMc}, both mass loss theories predict a rather sharp and peaked core mass distribution which are either shifted toward too high a mass \citep{Wu19} or too narrow \citep{Gupta19} with respect to the radial velocity follow-up of {\it Kepler} planets \citep{Marcy14}, even without correcting for a detection bias against small planets. \citet{Ginzburg18} found the 1-dimensional radius distribution expected from core-powered envelope mass loss to be stable between a core mass distribution that is flat below 5$M_\oplus$ and the Rayleigh distribution at least down to $\sim$1.3$R_\oplus$. \citet{Rogers21_pevap} obtain a much broader best-fit core mass distribution based on photoevaporation by using Bernstein polynomials to search for a non-parametric form of the distribution. They still expect a fall-off in the sub-Earth regime which is hard to reconcile with the lack of fall-off in the radius distribution at $\lesssim$1$R_\oplus$ in \citet{Hsu19}, see also Figure \ref{fig:obs_rdistrb}.

In this paper, we build on the work of \citet{Lee21} to investigate the role gas accretion physics plays in shaping the radius distribution of exoplanets from sub-Earths ($<1R_\oplus$) to giants ($\sim$10$R_\oplus$). Our work improves on the previous work in several important ways. First, we employ a more physically-motivated disk temperature profile as the amount of gas mass locked into the maximal isothermal envelope is sensitive to the nebular temperature. Second, we compare our theory to the occurrence rate studies of \citet{Hsu19} instead of \citet{Fulton18} as the former's approximate bayesian computation provides more accurate estimates near the sensitivity limit, which is the parameter range of our interest as we want to reproduce small planets at long orbital periods while also creating the radius gap. Third, we consider the entire radius distribution from those smaller than the Earth to those as large as Jupiter and including the radius gap to seek a unified theory that shapes the exoplanet size distribution.  

Our paper is organized as follows. In Section \ref{sec:methods}, we describe the model setup deriving the underlying disk profile, gas accretion, and how we convert envelope mass fraction into planet radii. Results are presented in Section \ref{sec:results} where we explore how our model shapes the radius distribution, the radius-period distribution, and the radius-stellar mass relation. We discuss the effect of post-formation mass loss as well as avenues for future studies in Section \ref{sec:disc}. Finally, we conclude and provide potential observational tests in Section \ref{sec:concl}.

\section{Methods}
\label{sec:methods}

\subsection{Onset of Gas Accretion}
\label{ssec:onset_accr}

We first establish the time at which we expect the gas accretion to begin in earnest. In the inner orbits of our interest ($\lesssim$300 days), the initial core assembly is extremely rapid. As long as the disk contains enough mass, the time it takes to build up a 5$M_\oplus$ core at $\sim$0.1 AU by planetesimal accretion is at maximum 0.01 Myrs \citep[e.g., see][their equations 1 \& 2]{Lee14}, which is orders of magnitude shorter than the typical disk lifetime of $\sim$1--10 Myrs (\citealt{Mamajek09}; see also the updated longer disk lifetimes measured by \citealt{Michel21}). This coagulation timescale will only shorten if there is gravitational focussing \citep[e.g.,][]{Goldreich04} including the effect of pebble accretion \citep{Ormel10,Lambrechts14}. By contrast, gas accretion timescale is comparable to the disk lifetime and, more importantly, the rate of accretion increases with core mass \citep[e.g.,][]{Lee15} which implies that it is the last gas mass doubling that matters most when it comes to gas accretion \citep[see also][]{Piso14}. This separation in timescale between the initial build up of the core and gas accretion allows us to safely ignore the initial stage of core formation in our calculation.

What cannot be ignored is the later stage of core assembly process. The final masses of planetary cores are likely shaped by late-stage orbital instability and collisional mergers \citep[e.g.,][]{Kokubo98,Kominami02,Hansen12,Hansen13}, with the final orbital architecture and composition being sensitively determined by the initial solid distribution \citep[e.g.,][]{Dawson15,Moriarty16} and the dissipation history of the underlying disk gas \citep[e.g.,][]{Dawson16}. In fact, this late stage \textit{in situ} orbital instability is required to reproduce the fact that the majority of Kepler multi-planet systems are no where near resonance \cite[][see also \citealt{Swain19} for a hint of density enhancement by collisions]{Izidoro17}. In collisional mergers, it is the last doubling that takes the longest as the orbit-crossing timescale is exponentially dependent on orbital spacing \citep[e.g.,][]{Zhou07}. Furthermore, upon collision, the protocores can lose up to order unity amount of their gaseous envelope, depending on the impact angle and the impact velocity \citep[e.g.,][]{Inamdar16}. Combined with the fact that the rate of gas accretion is larger for more massive cores, it follows that the most amount of gas envelope is built after the core has completed its assembly process. 

By comparing the gas eccentricity damping time and the orbit crossing time, \citet{Lee16} estimated that the disk gas needs to be depleted by factors of at least $\sim$0.0005 with respect to the solar nebula \citep[see also][]{Kominami02}. Orbit crossing is a probabilistic event and the orbit crossing time quoted in the literature refers to the time at which 50\% of the simulated systems undergo instability \citep[see e.g.,][]{Pu15}. Using direct N-body integrations, \citet{Dawson16} found that orbit crossing can begin at milder gas depletion of $\sim$0.01 with respect to the solar nebula (for their chosen reference disk properties, this is equivalent to 3 orders of magnitude depletion with respect to the minimum mass extrasolar nebula (MMEN) computed by \citet{Chiang13}). Gas depletion by 3--5 orders of magnitude with respect to the solar-composition nebula is also favored for the reproduction of the orbital period ratio distribution of multi-planetary systems using the small amount of eccentricity damping and migration in the gas-poor environment \citep{Choksi20}, and so we choose the time at which the disk gas depletes by this range of values at each orbital period as the onset time for gas accretion. 

\subsection{Disk Temperature}
\label{sse:diskT}

The isothermal maximum envelope mass is exponentially sensitive to the disk temperature at the location of the planetary core. \citet{Lee21} fixed the overall temperature profile to that of passive disks and let the normalization be a free parameter. Here, we derive more realistic disk temperature profiles.

First, we determine whether the disk is optically thin or thick to the incoming stellar light and outgoing thermal radiation. We compute the vertical optical depth through the disk to the incoming stellar irradiation and outgoing internal thermal radiation by integrating density and temperature-dependent Rosseland mean opacity over three vertical scale heights (the disk is assumed to be locally and vertically isothermal for simplicity). The opacities are calculated using the methodology outlined in Section 2.2 of \citet{Lee18}. To summarize, we compute the Rosseland mean opacity to stellar irradiation $\kappa_\star$ and internal thermal radiation $\kappa_{\rm th}$ by integrating wavelength-specific opacities $\kappa$ generated from a modified version of the stellar atmosphere code \texttt{PHOENIX} as described in \citet{Ferguson05}, assuming solar metallicity (Z = 0.02, where elemental abundances are scaled to those in \citet{GN93}). In `dusty' models, metals take the form of dust grains following interstellar medium (ISM) size distribution whereas in `dust-free' models, dust grains do not contribute to the opacity (due to e.g., coagulation); any metallic contribution to the opacity is from gaseous forms. Our choice of \citet{Ferguson05} over \citet{Freedman14} is motivated by the former's inclusion of heavy atomic metals such as iron for more completeness. In calculating $\kappa_\star$, the Planck function is evaluated at the stellar effective temperature (taken at 4000 K to be representative of pre-main sequence stars) whereas for $\kappa_{\rm th}$, the Planck function is evaluated at the local temperature.\footnote{The vertical optical depth to stellar irradiation barely changes between stellar effective temperatures of 3000 and 5000 K, relevant for stars of masses $\sim$0.1--1.2$M_\odot$ in their pre-main sequence phase.} Optical depths are computed by integrating $\kappa_\star$ and $\kappa_{th}$ vertically over 3 scale heights for vertically isothermal disks over a wide range of midplane density.

\begin{figure}
    \centering
    \includegraphics[width=0.5\textwidth]{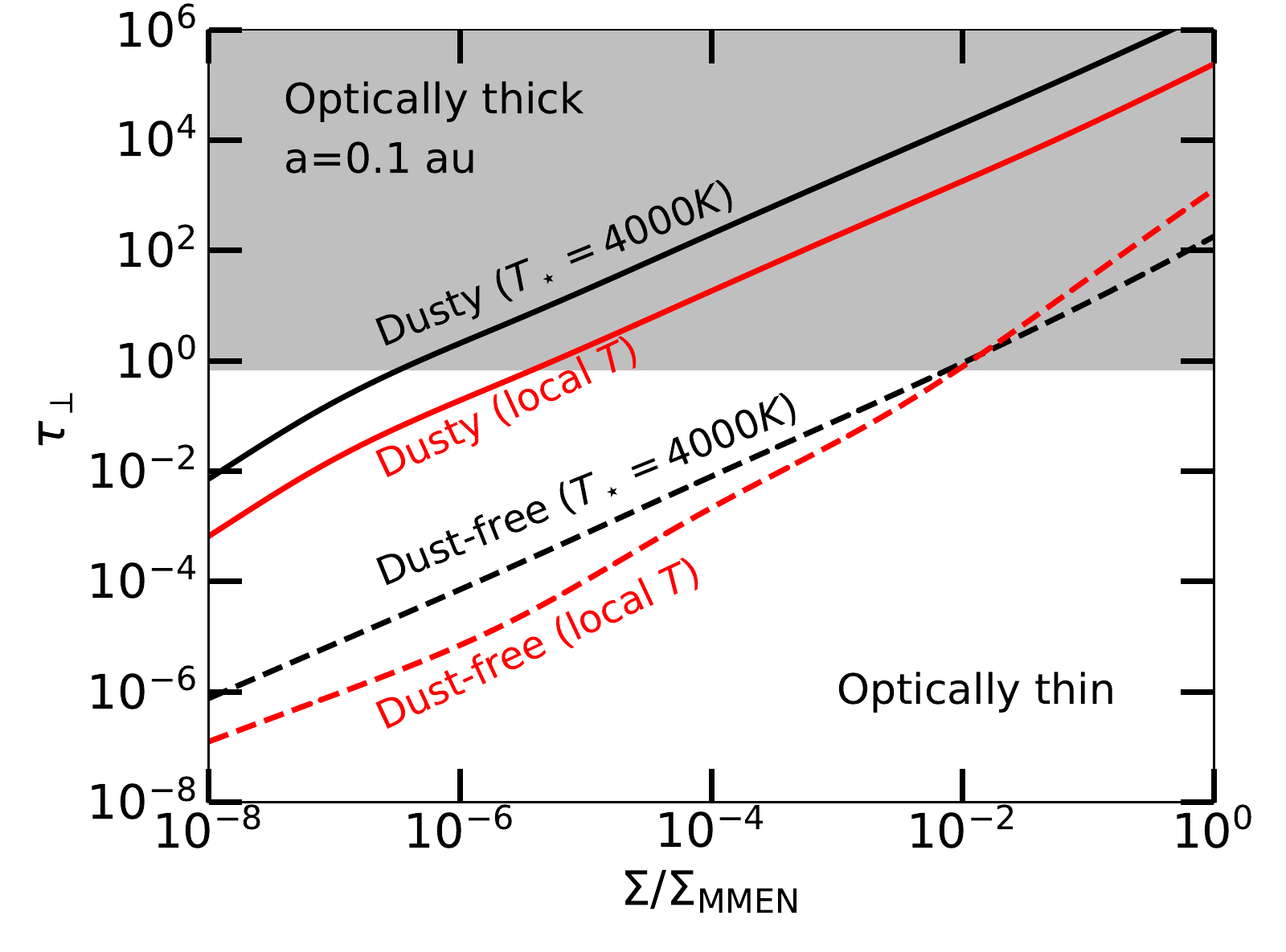}
    \caption{Vertical optical depth to stellar irradiation (black) and internal thermal radiation (red) vs.~disk gas surface density, normalized to that of minimum mass extrasolar nebula (MMEN; \citealt{Chiang13}). When dust grains following the ISM grain size distribution dominate the opacity (solid lines), the disk remains optically thick at 0.1 au until the gas depletes by more than 6 orders of magnitude with respect to MMEN. When dust grains do not contribute to opacity (dashed lines), the disk becomes optically thin once the gas depletes by approximately 2 orders of magnitude.}
    \label{fig:optdepth_rho}
\end{figure}

Figure \ref{fig:optdepth_rho} demonstrates that disks depleted by more than 2 orders of magnitude with respect to the minimum mass extrasolar nebula will be optically thin to both incoming stellar irradiation and outgoing internal thermal radiation when dust grains do not contribute to the overall opacity (`dust-free'). When dust grains are the dominant source of opacity, however, the disk remains optically thick down to $\Sigma/\Sigma_{\rm MMEN} \sim 10^{-6}$ (for internal radiation) and $\sim$10$^{-7}$ (for stellar irradiation), where $\Sigma$ is disk gas surface density and $\Sigma_{\rm MMEN}$ is the gas surface density of minimum mass extrasolar nebula \citep{Chiang13}. 

It follows that over the range of disk gas depletion required for the final core assembly, a `dust-free' disk presents an optically thin environment. As we will show below, being optically thin down to the midplane simplifies considerably the calculation of temperature profile. \citet{Lee21} favored dust-free accretion in order to explain the radius gap from primordial gas accretion and so we focus our attention to dust-free case in this paper and revisit the dusty case in future work with a brief discussion in Section \ref{ssec:dusty_v_df}. 

In general, the amount of total flux received from stellar irradiation dominates the dissipation flux from viscous accretion, and so if the disk is optically thin down to the midplane to both incoming and outgoing radiation, the midplane temperature is regulated by irradiation.
We consider the trace amount of individual dust grains everywhere in the disk are being directly heated by the starlight so that they are heated to temperatures of
\begin{equation}
    T_d^4 = \frac{L_\star (t)}{16\pi \sigma_{\rm sb} a^2 \epsilon}
\end{equation}
where $L_\star (t)$ is the luminosity of the central star that varies with time as the star evolves (we describe more at the end of this section how we account for stellar evolution), $\sigma_{\rm sb}$ is the Stefan-Boltzmann constant, $a$ is the orbital distance, and $\epsilon$ is the emissivity of the grains, $\epsilon \propto T_d^\beta$ \citep[e.g.,][]{Chiang97,Garaud07}. We adopt the relevant temperature profile estimated by \citet{Chiang97}:
\begin{equation}
    T_d = 550\,{\rm K}\,\left(\frac{a}{1 {\rm au}}\right)^{-2/5}\left(\frac{L_\star (t)}{5.6\times 10^{33}\,{\rm erg/s}}\right)^{1/5}
    \label{eq:Tdisk}
\end{equation}
which assumes $\beta=1$; yet, it is a close approximation of the more accurate solution found by \citet[][see their Figure 5]{Chiang01}. Since gas molecules are poor absorbers and emitters of optical and near infrared photons, we consider gas to be thermally equilibrated to the dust grains so that the disk temperature is set to $T_d$ written in equation \ref{eq:Tdisk}. The fact that our disk is optically thin down to the midplane begets a disk temperature that scales differently from the passive, but optically thick disk, as was adopted by \citet{Lee21}.

For stellar luminosities, we adopt the \citet{Johnstone21} stellar evolution track which takes into account the full pre-main sequence stage evolution and the effect of stellar spin. For all our calculation, we adopt their median spin for a given stellar mass which derives from the measured rotation distributions of nearby young clusters. 

With these setups, our disk temperature is $\lesssim$1000 K over the range of orbital distances of our interest, which is colder than the favored solution of \citet{Lee21}. As we will show below, even with this colder but more accurate, disk temperature, gas accretion physics can reproduce the observed radius gap.

\subsection{Disk Gas Surface Density and Radial Accretion}
\label{ssec:gas_sdens}

We adopt the viscously evolving disk undergoing photoevaporation to model the late stages of disk evolution (e.g., transitional disks; \citealt{Owen11}; see also \citealt{Alexander14} and \citealt{Pascucci22} for reviews). Viscous + photoevaporation has been shown to explain roughly half of the observed transitional disks (i.e., those with accretion rates below $10^{-9}\,M_\odot\,{\rm yr}^{-1}$ and cavities out to $\sim$30 au; \citealt{Picogna19}). Those with higher accretion rates are likely explained by magnetized disk winds \citep[e.g.,][]{Suzuki16,Wang17} whose exploration in the context of exoplanetary radius distribution is beyond the scope of this paper and we consider them in future work.

As we have already established that most gas accretion occurs after the final assembly of cores in gas-depleted environment (Section \ref{ssec:onset_accr}), we focus our attention after the photoevaporative winds have carved out a gap at a few AU and the inner disk has been decoupled from the outer disk so that the former drains out on a viscous timescale set at the truncation radius $\sim$3 AU \citep[see e.g.,][their Figure 9]{Owen11}. We model the rapid drop in surface density in the inner orbits as an exponential fall-off in time and retain the spatial dependence of orbital distance as expected from the similarity solution to disk momentum equation under viscous torque as derived by \citet{Lynden-Bell74} and \citet{Hartmann98}:
\begin{equation}
        \frac{\Sigma_{\rm bg}(a, \Delta t)}{\rm g\, cm^{-2}} = 10^4 f_{\rm dep,1} \left(\frac{a}{1\,{\rm AU}}\right)^{\frac{-11}{10}}\exp\left[-\frac{\Delta t}{t_{\rm visc}}\right]
    \label{eq:Sigma_gas1}
\end{equation}
where $f_{\rm dep,1}=0.01$ is the initial depletion factor with respect to MMEN (i.e., at the time of disk truncation by winds; we adopt 0.01 by visual inspection of Figure 9 of \citealt{Owen11}), $\Delta t$ is time since the truncation of the disk by winds (we set the truncation time at 4 Myrs), and $t_{\rm visc}$ is the characteristic draining time. The power-law scaling on $a$ derives from $\Sigma_{\rm bg} \propto a^{-\gamma}$ where the kinematic viscosity $\nu \propto a^{\gamma}$ with $\nu = \alpha c_s H$, $c_s = \sqrt{kT_d/\mu m_H}$ is the sound speed, $k$ is the Boltzmann constant, $\mu=2.34$ is the mean molecular weight, $m_H$ is the mass of atomic hydrogen, $H = c_s/\Omega$ is the disk scale height, $\Omega = \sqrt{G M_\star/a^3}$ is the Keplerian orbital frequency, $G$ is gravitational cosntant, and $M_\star$ is the mass of the host star. The Shakura-Sunyaev parameter $\alpha$ is taken as spatially constant. For a disk accretion rate of $\dot{M}_{\rm disk} = 3\pi\nu\Sigma_{\rm bg} = 10^{-9} M_\odot\,{\rm yr}^{-1}$, $\nu \sim 10^{14} {\rm cm}^2\,{\rm s}^{-1}$ at the disk truncation radius 3 AU and $\Delta t = 0$. Then, $t_{\rm visc}$ evaluated at the truncation radius 3 AU is $\sim$0.45 Myrs and so we set $t_{\rm visc}=0.45$ Myrs, and the time-evolving disk accretion rate is

\begin{equation}
    \dot{M}_{\rm disk} = 10^{-9}M_\odot\,{\rm yr}^{-1}\exp\left[-\frac{\Delta t}{t_{\rm visc}}\right].
    \label{eq:Mdot_disk}
\end{equation}

\begin{figure*}
    \centering
    \includegraphics[width=\textwidth]{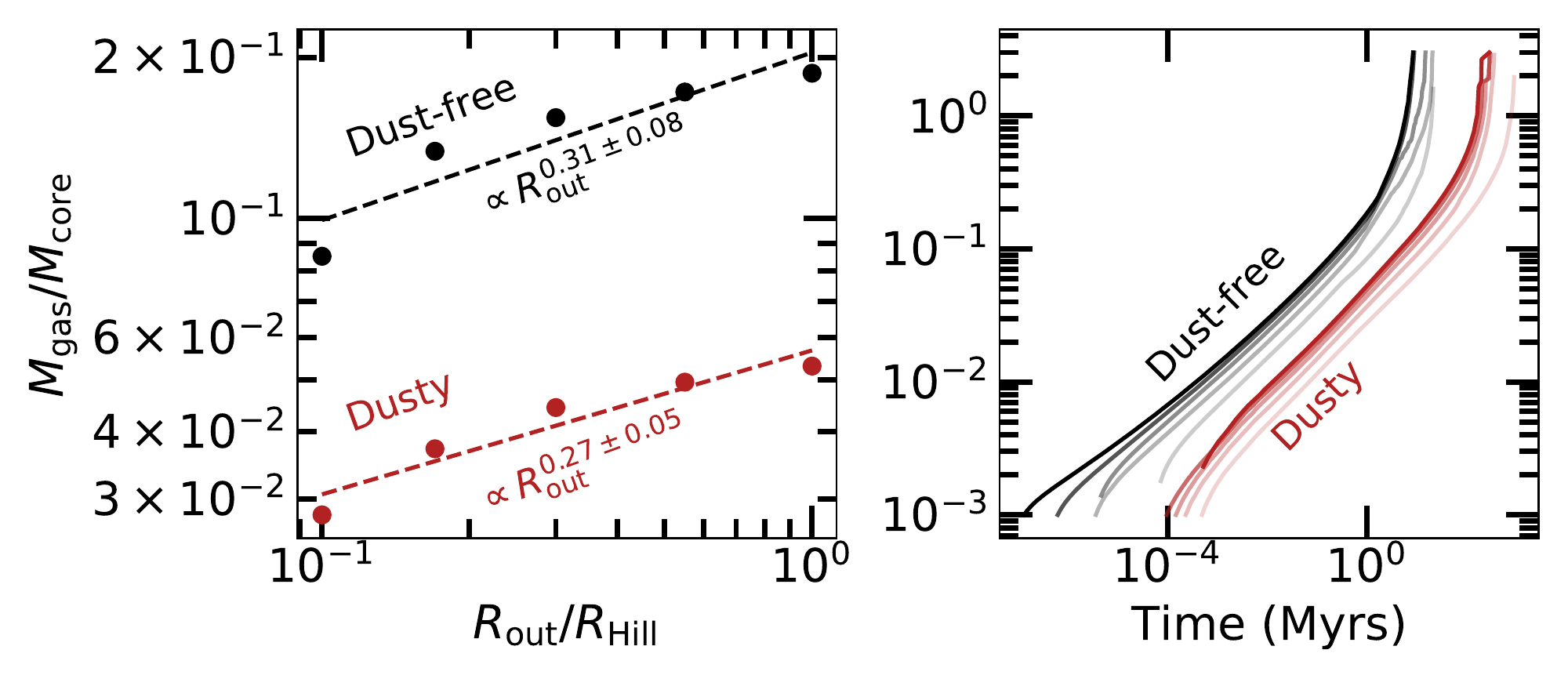}
    \caption{Envelope mass fraction of a 5 $M_\oplus$ core at 0.3 au in MMEN, with its gas surface density reduced by a factor of 0.01. Left: how the final $M_{\rm gas}/M_{\rm core}$ at 1 Myr scales with the outer boundary radius. We observe a weak dependence on the bound radius down to $\sim$0.1$R_{\rm out}/R_{\rm Hill}$. Right: time evolution of $M_{\rm gas}/M_{\rm core}$ with lighter curves representing smaller $R_{\rm out}/R_{\rm Hill} =(1.0,0.55,0.3,0.17,0.1)$, as shown on the left panel.}
    \label{fig:gcr_v_t_rscl}
\end{figure*}

\begin{figure}
    \centering
    \includegraphics[width=0.5\textwidth]{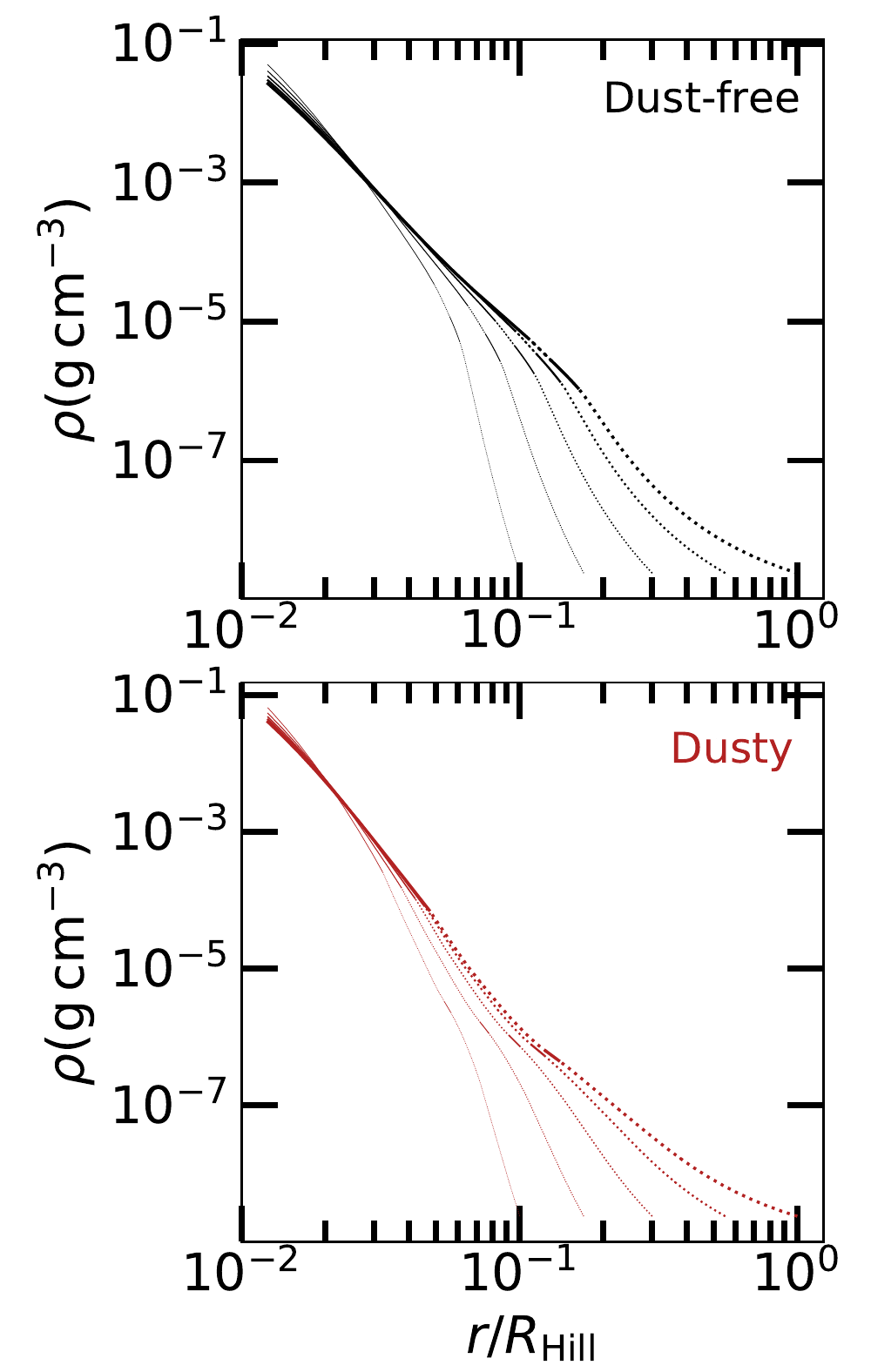}
    \caption{Radial density profiles of a 5$M_\oplus$ core at 0.3 au with $M_{\rm gas}/M_{\rm core} = 0.01$. The underlying disk is MMEN depleted by a factor of 100. Like Figure \ref{fig:gcr_v_t_rscl}, lighter curves represent smaller $R_{\rm out}/R_{\rm Hill}=(1.0,0.55,0.3,0.17,0.1)$ while solid and dotted lines illustrate convective and radiative zones, respectively. The small convective windows are created by a small increase in opacities at temperatures of $\sim$800 K, likely corresponding to the water absoprtion line at $\sim$3$\mu$m. Smaller $R_{\rm out}$ steepens the overall density profile, rendering the radiative-convective boundary (rcb) more optically thick, although these differences are minimal down to $R_{\rm out}/R_{\rm Hill}=0.1$.}
    \label{fig:atmprof_rscl}
\end{figure}

\subsection{Gas Accretion}
\label{ssec:accr_cool}

The rate at which a core accretes its gaseous envelope can be regulated by either cooling or hydrodynamic delivery \citep[e.g.,][]{Pollack96,Tanigawa16}. We largely adopt the methodology of \citet{Lee19} in computing the envelope mass accreted onto our planet except we scale from the full numerical solution of \citet{Lee14} and \citet{Lee16} instead of approximating the runaway as an exponential growth. 
As the goal of this paper is to assess the entire radius distribution up to $\sim$10$R_\oplus$, we need to properly account for the runaway growth. Here, we summarize the key points while highlighting the differences.

\subsubsection{When Gas Accretion Begins and Ends}
In Section \ref{ssec:onset_accr}, we established that the final core assembly occurs when the disk is depleted by 3--5 orders of magnitude with respect to MMEN. Parameterizing the target depletion factor as $f_{\rm dep}$, at each orbital distance, we begin gas accretion at:
\begin{equation}
    t_{\rm start} = - t_{\rm visc}\log\left[\left(\frac{f_{\rm dep}}{f_{\rm dep,1}}\right)a^{-1.6+11/10}\right].
    \label{eq:tstart}
\end{equation}
\citet{Lee18} found that the accretion by cooling halts when the disk depletes by more than 8 orders of magnitude with respect to MMEN as the ambient density is so low it forces the entire envelope to become radiative, and so we stop the accretion at
\begin{equation}
    t_{\rm end} = - t_{\rm visc}\log\left[\left(\frac{10^{-8}}{f_{\rm dep,1}}\right)a^{-1.6+11/10}\right].
    \label{eq:tend}
\end{equation}
Note that at such severe depletion, even the global disk accretion would be minimal and so it is safe to consider all gas accretion to have halted.

\subsubsection{Accretion by Cooling}

We integrate the stellar structure equations 5--8 of \citet{Lee14} from the bound radius of the envelope $R_{\rm out} = f_R {\rm min}(R_{\rm Hill}, R_{\rm Bondi})$---where $f_R$ is a numerical factor, $R_{\rm Hill}$ the Hill radius, and $R_{\rm Bondi}$ the Bondi radius---down to the inner boundary set at the core radius $R_{\rm core}/R_\oplus = f_{\rho_c}^{-1/3}(M_{\rm core}/M_\oplus)^{1/4}$ \citep{Valencia06}, where $f_{\rho_c}$ a numerical factor of core density with respect to that of the Earth, and $M_{\rm core}$ is the mass of the core. Compared to $R_{\rm core} \propto M_{\rm core}^{1/3}$ used by \citet{Lee14}, this makes negligible difference in the envelope mass fraction but could lead to observable difference when translated to radius so we adopt the $R_{\rm core} \propto M_{\rm core}^{1/4}$ scaling for more accuracy. At $R_{\rm out}$, the density and the temperature of the planet smoothly connects to that of the disk. We use the opacity tables built by the methodologies of \citet{Ferguson05} and use the equation of state of the gaseous envelope composed of hydrogen, helium, and metallic species as computed in \citet{Lee14}.

Three-dimensional hydrodynamic calculations report $f_R$ can be as low as 0.1--0.3 \citep{Lambrechts17,Fung19} as the advective flow of the disk can shrink the extent of the bound trajectory of gas around the planet.
While some hydrodynamic simulations report complete recycling of gas or the advection alone completely limiting the envelope growth beyond a few percent by mass \citep[e.g.,][]{Bethune19,Moldenhauer21}, these calculations are limited by their difficulty in resolving the innermost convective envelope whose mass is centrally concentrated \citep{Lee14} and therefore more likely to be shielded from the effects of the outer flow. 
Figure \ref{fig:gcr_v_t_rscl} demonstrates that the rate of envelope mass growth decreases with smaller bound radius but its effect is weak. First, this decrease can be explained by the steeper density profile and therefore more optically thick radiative-convective boundary (rcb) for smaller $R_{\rm out}$. As Figure \ref{fig:atmprof_rscl} illustrates, to pack the same amount of envelope within a smaller spatial extent, the density profile necessarily steepens and so the density and therefore the optical depth at the rcb rises. Since the rcb regulates the rate of cooling and therefore accretion, slower cooling begets slower accretion. The change in the density at the rcb however is only of order unity as long as $R_{\rm out}$ remains larger than the initial location of the rcb which is $\sim$0.1$R_{\rm Hill}$. Even at $R_{\rm out}/R_{\rm Hill} = 0.1$, the envelope mass fraction still drops by factors of $\lesssim$2.

\citet{Ali-Dib20} showed that in disks with sufficiently high entropy (equivalent to high temperature and low density), advection of entropy into the deeper layers of the envelope can stall its mass growth by an erasure of the inner convective zone. Such behavior can also be explained from the perspective of density steepening. For a fixed envelope mass, smaller disk gas density requires a steeper envelope density profile. This requirement is exacerbated by the entropy advection which forces the outermost layer to be convective (i.e., shallower density profile) and only becomes worse if the penetration depth of the advective flow is assumed to be deeper. For high enough envelope mass, the solution to structure equations demands that the innermost profiles to be steep and therefore fully radiative. In our calculations, we always find the innermost convective zone except when the disk gas surface density drops by more than 8 orders of magnitude \citep{Lee18}. Recently, \citet{Zhu_zhaohuan21} performed detailed hydrodynamic simulations of an accreting planet embedded in a disk with realistic opacity to show that the recycling flow has limited effect on the envelope thermal structure and that the upper envelope never becomes isentropic but rather looks identical to an isolated envelope. Their calculation was benchmarked to Jupiter at 5 AU so an analogous calculation at closer-in orbits would be welcome to verify if their results scale to short distances. For this study, we adopt $f_R = 0.2$ throughout.

\subsubsection{Scaling to Different Cores and Disk Conditions}
\label{sssec:scaling_cool}

We use the analytic scaling relations found by \citet{Lee15} to scale our numerical calculations of $M_{\rm gas}/M_{\rm core}$ to a wide variety of core and disk properties. (While we show both dusty and dust-free scaling here for completeness, we will focus on dust-free accretion for computing the expected radius distribution.) The scaling indices differ slightly due to our use of $R_{\rm core} \propto M_{\rm core}^{1/4}$ instead of $R_{\rm core} \propto M_{\rm core}^{1/3}$. Since what changes is the rate of gas accretion and more specifically the cooling timescale, we apply scaling to the time, rather than $M_{\rm gas}/M_{\rm core}$. More technically speaking, we shift the time axis for our base $M_{\rm gas}/M_{\rm core}(t)$ computed numerically for 5$M_\oplus$ core at 0.3 au ($\rho_{\rm disk}=2.4\times 10^{-9}\,{\rm g\,cm^{-3}}, T_{\rm disk}=600\,{\rm K}$, $R_{\rm out}/R_{\rm Hill}=0.55$ for dust-free and 0.17 for dusty):
\begin{align}
    M_{\rm gas}/M_{\rm core} &\propto t^{0.4} f_R^{0.31} T_d^{-1.5} Z^{-0.4} \mu_{\rm env}^{2.2} M_{\rm core}^{1.8} \Sigma_{\rm bg}^{0.12} \nonumber \\ 
    t_{\rm scaled} &\propto t_{\rm base} f_R^{-0.77} T_d^{3.75} Z \mu_{\rm env}^{-5.5} M_{\rm core}^{-4.5} \Sigma_{\rm bg}^{-0.3},
    \label{eq:gcr_tscl_df}
\end{align}
for dust-free, and
\begin{align}
    M_{\rm gas}/M_{\rm core} &\propto t^{0.4} f_R^{0.28} Z^{-0.4} \mu_{\rm env}^{3.4} M_{\rm core}^{1.8} \Sigma_{\rm bg}^{0.12} \nonumber \\ 
    t_{\rm scaled} &\propto t_{\rm base} f_R^{-0.70} Z \mu_{\rm env}^{-8.5} M_{\rm core}^{-4.5} \Sigma_{\rm bg}^{-0.3},
\end{align}
for dusty, where $Z$ is envelope metallicity and $\mu_{\rm env}$ is the mean molecular weight of the envelope. We solve for $\mu_{\rm env}$ self-consistently from an input $Z$ which is set to solar metallicity 0.02 throughout this paper. The strong sensitivity of $t_{\rm scaled}$ on $\mu_{\rm env}$ is only relevant for high $Z \gtrsim 0.2$--0.4 \citep[see][their Figure 1]{Lee16}. 

With the dependence on $\mu_{\rm env}$ being largely irrelevant, the rate of gas accretion is most sensitive to $M_{\rm core}$ followed by $T_{\rm d}$ under dust-free accretion. This strong sensitivity on $M_{\rm core}$ allows us to constrain the underlying core mass distribution of exoplanets using the gas accretion physics, and as we explain in Section \ref{ssec:mass_time_per}, it is one of the main free parameters in our calculation. Likewise, the strong sensitivity to $T_{\rm d}$ motivates the more careful calculation of the disk temperature which is one of the main updates of this paper compared to previous calculations.

We note that the dust-free $M_{\rm gas}/M_{\rm core}$ scales more strongly with $M_{\rm core}$ than what was used by \citet{Lee21}. Such difference stems from the latter's choice of using the ``dust-free and gas-rich beyond 1 au'' relation of \citet{Lee15}. We have verified numerically that $M_{\rm gas}/M_{\rm core} \propto M_{\rm core}^{1.8}$ is more correct in the parameter space of our interest (inside 1 au and gas-depleted). When computing the rate of gas accretion by cooling $\dot{M}_{\rm cool}$, we numerically differentiate the base calculations, apply a Savitzky-Golay smoothing filter (from \texttt{SciPy.signal}) to avoid discontinuity due to numerical resolution, then interpolate the smoothed $\dot{M}_{\rm cool}$ with respect to $t_{\rm scaled}$ appropriately scaled for a given core and a given set of disk properties.

\subsubsection{Hydrodynamic Limit for Post-Runaway Planets}

Once $M_{\rm gas}/M_{\rm core} \gtrsim 0.5$, accretion by cooling enters the runaway phase triggered by thermal disequilibrium: large cooling power is required to keep massive envelopes in hydrostatic equilibrium which accelerates gas accretion requiring even larger cooling power, catastrophically shortening the cooling timescale. Once the runaway accretion is triggered, the rate of gas accretion is limited not by thermodynamics but rather by hydrodynamic delivery \citep[e.g.,][]{Pollack96} whether by local delivery or by global disk accretion, whichever provides smaller accretion rate. We show below that the global disk accretion limits the post-runaway accretion at all times for the parameters of our interest.

From equations 7 and 8 of \citet{Tanigawa16}, the rate of local gas delivery is
\begin{equation}
\dot{M}_{\rm hydro} = 0.29 \left(\frac{M_{\rm p}}{M_\star}\right)^{4/3} \Sigma_{\rm neb} \left(\frac{a}{H}\right)^2 a^2 \Omega.
\label{eq:Mdot_hydro_norm}
\end{equation}
where $M_p = M_{\rm core} + M_{\rm gas}$ is the total planet mass, and $\Sigma_{\rm neb} = \Sigma_{\rm bg}/(1+0.034K)$ with
\begin{equation}
    K = \left(\frac{H}{a}\right)^{-3}\left(\frac{M_{\rm p}}{M_\star}\right)^2\left(\frac{\Omega a^2}{\nu}\right).
    \label{eq:K}
\end{equation}
The depletion factor $K$ represents the gap opening by the one-sided Lindblad torque of a planet against the viscous torque of the disk \citep[e.g.,][]{Duffell13,Fung14}. The scaling relationship of equation \ref{eq:Mdot_hydro_norm} can be understood as the shock of accretional flow at the planet-disk interface for $R_{\rm Hill} > H$ (equivalently, $R_{\rm Hill} < R_{\rm Bondi}$).\footnote{This particular scaling is relevant for isothermal shocks. We note that for the parameters of our interest (inner disk that is gas-depleted), the shock cooling timescale is significantly shorter than the orbital time so that the isothermal shock approximation is valid \citep[see][their equation 8]{Lee19}. We also note that planets that trigger runaway are massive enough that their $R_{\rm Hill} > H$ (i.e., all our planets that enter post-runaway accretion are super-thermal in contrast to sub-thermal planets at wide orbits studied by \citet{Ginzburg19}).}

At the fiducial $R_{\rm out}/R_{\rm Hill} = 0.2$ in the inner orbits, the cores need to be $\gtrsim 10M_\oplus$ to enter the post-runaway phase before the disk gas dissipates away. This mass is either greater than or around the minimum total planet mass required to open a deep gap ($0.034 K > 1$) at the orbital distances of our interest which is
\begin{equation}
    M_{\rm p,gap} \sim 13 M_\oplus \left(\frac{a}{1\,{\rm AU}}\right)^{3/4}.
\end{equation}
In the limit of deep gap ($\Sigma_{\rm neb} \sim \Sigma_{\rm bg}/0.034 K$), we calculate the minimum total mass of the planet needed for the local hydrodynamic delivery to be the limiting process over the global disk accretion (here written as $\dot{M}_{\rm disk} = 3\pi\nu\Sigma_{\rm bg}$):
\begin{align}
    \frac{\dot{M}_{\rm hydro}}{\dot{M}_{\rm disk}} &= \frac{0.29}{3\pi}(0.034)^{-1}\left(\frac{M_p}{M_\star}\right)^{-2/3}\left(\frac{H}{a}\right) < 1 \nonumber \\
    M_{\rm p,hyd} &> 8.2 M_{\rm jup} \left(\frac{a}{1\,{\rm AU}}\right)^{9/20},
\end{align}
which implies $\dot{M}_{\rm hydro} < \dot{M}_{\rm disk}$ only for super-Jupiters. Using equation \ref{eq:Mdot_disk}, we find that our planets can at best reach $\sim$0.3$M_{\rm jup}$ for our fiducial parameters and so in our gas accretion calculation, we only check for whether $\dot{M}_{\rm cool} > \dot{M}_{\rm disk}$ which we define as post-runaway accretion. We are implicitly assuming all the gas that diffuses past the planet's orbit is accreted by the planet as found by hydrodynamic simulations \citep[e.g.,][]{Lubow06}. Our finding that $\dot{M}_{\rm disk} < \dot{M}_{\rm hydro}$ always once the hydrodynamic limit is reached agrees with \citet{Rosenthal20} who found the same result in the outer disk undergoing viscous accretion.

\subsubsection{Isothermal Limit}

All planetary cores at a given mass and given location in a disk have maximum possible envelope mass they can accrete, given by the maximally cooled isothermal limit (i.e., when the entire envelope thermally relaxes to the outer nebula and reach a full isothermal profile):
\begin{equation}
    \label{eq:Miso}
    M_{\rm iso} = 4\pi\rho_{\rm disk}\int^{R_{\rm out}}_{R_{\rm core}} r^2\,{\rm Exp}\left[\frac{GM_{\rm core}}{c_{s}^2}\left(\frac{1}{r}-\frac{1}{R_{\rm out}}\right)\right]dr,
\end{equation}
where $\rho_{\rm disk} = \Sigma_{\rm bg}/H$ is the volumetric disk gas density. 

For small cores ($\lesssim$1--2$M_\oplus$), this isothermal limit is so small that they could never have been sub-Neptunes to begin with. The steep exponential drop in the expected envelope mass fraction for these planets is the reason why we expect the radius gap to be set in place primordially.

\subsection{Core Mass and Orbital Period}
\label{ssec:mass_time_per}

How much envelope a planet ends up with depends sensitively on how massive their cores are and when the cores assemble \citep{Lee19}. From the radial-velocity follow-up of {\it Kepler} planets by \citet{Marcy14}, we infer that the underlying core mass distribution should drop beyond $M_{\rm core} \gtrsim$4--6$M_\oplus$ (see Figure \ref{fig:dNdMc} where we reproduce the results of \citet{Marcy14}). \citet{Otegi20} also report a fall-off in the number of planets with masses beyond $\sim$5--10$M_\oplus$ (see their Figure 1) where they select planets with precise radii and mass measurements from an amalgam of different surveys collated from the NASA Exoplanet Archive. Whether the distribution falls at lower masses or not is unclear at this point due to the difficulty in measuring low masses. 

The bottom-heavy radius distribution of \citet{Hsu19} for $R \lesssim 2R_\oplus$ suggests a potentially bottom-heavy core mass distribution. From the radial velocity follow-ups, we see no sub-Neptunes or super-Earths more massive than $\sim$20$M_\oplus$; furthermore, larger sub-Saturns and giants are approximately an order of magnitude less numerous than their smaller counterparts \citep[e.g.,][]{Howard10,fressin13,petigura13}. Motivated by these observations, we assume a following core mass distribution
\begin{equation}
    \frac{dN}{d\log M_{\rm core}} \propto M_{\rm core}^{\delta} \exp\left(-\frac{M_{\rm core}}{M_{\rm break}}\right).
    \label{eq:dNdlogMc}
\end{equation}
From our experiment and visual inspection, we find $\delta=0.3$ (top-heavy) and $\delta=-0.3$ (bottom-heavy) with $M_{\rm break}=4M_\oplus$ show a good agreement with the observations and so we focus on those parameters to demonstrate the viability of primordial gas accretion in reproducing the observed radius distribution for a broad (and even bottom-heavy) core mass distribution in more realistic disk environment.

The occurrence rate of sub-Neptunes remains flat down to $\sim$10 days inside which they drop \citep{Youdin11,fressin13,Petigura18}. The break at $\sim$10 days can be explained by the magnetospheric truncation of the underlying disk whose inner radii are expected to be at co-rotation radius with respect to their host star spin \citep{Mulders15,Lee17}. We draw the planet orbital periods $P$ from 0.5 to 500 days following the distribution
\begin{equation}
    \frac{dN}{d\log P} \propto 1-\exp\left[-\left(\frac{P}{11.9\,{\rm days}}\right)^{2.4}\right]
    \label{eq:dNdlogP}
\end{equation}
which follows closely the fitting result of \citet{Petigura18} with a modification to enforce a log-uniform distribution of $P$ beyond $\sim$10 days. The flat $P$ distribution ensures that any dependence of planet size with orbital period (beyond 10 days) we see in our model is from gas accretion physics and not from an a priori assumption of planet orbital periods. Although planets larger than Neptune are observed to follow a different orbital period distribution, we adopt the above distribution uniformly across all our planets for simplicity.

\subsubsection{Implementation}

We follow the steps outlined below to compute the envelope mass fraction:

\begin{enumerate}
    \item For each core, check if the envelope radius is larger than the core radius ($R_{\rm out} = f_R {\rm min}(R_{\rm Hill}, R_{\rm Bondi}) > R_{\rm core}$). If not, we stop the calculation and give zero envelope mass to the planet (physically, this planet is unable to hold on to a bound envelope). Otherwise we proceed.
    
    \item For each core with a bound envelope, we compute the maximal isothermal mass following equation \ref{eq:Miso}.
    
    \item We numerically solve the differential equation $dM_{\rm gas}/dt = {\rm min}(\dot{M}_{\rm cool}, \dot{M}_{\rm disk})$ from $t_{\rm start}$ to $t_{\rm end}$ at logarithmic time interval of $\log t = 0.001$ (equivalent to $\gtrsim$1000 time steps) using a simple leapfrog scheme.
\end{enumerate}

\begin{figure}
    \centering
    \includegraphics[width=0.5\textwidth]{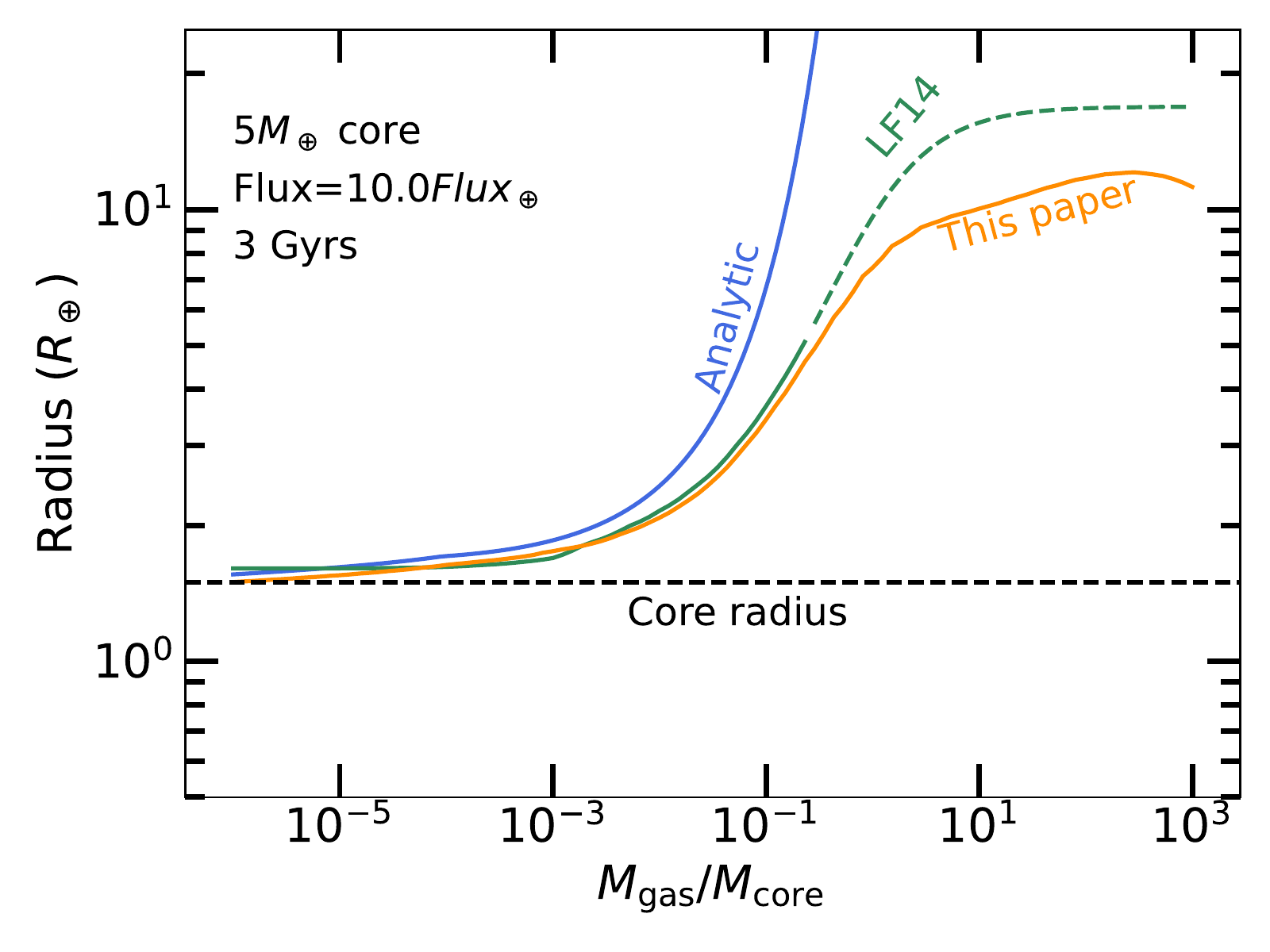}
    \caption{Comparison of radius conversion methods between the analytic scheme of \citet[][blue solid]{Lee21}, extrapolating the grid of \citet[][green; LF14]{Lopez14} and the grid presented in this paper (orange solid). The analytic scheme does not take into account the self-gravity of the envelope and therefore overestimates the radius at $M_{\rm gas}/M_{\rm core} \gtrsim 0.1$. LF14 grid only extends up to $M_{\rm gas}/M_{\rm core} \sim 0.25$ (region covered by the grid plotted in solid; extrapolation in dashed). Using the radius calculation presented in this work, we see a slight downturn in the radius at $M_{\rm gas}/M_{\rm core} \sim 300$ (equivalent to $\sim$5$M_{\rm jup}$) as we approach the brown dwarf regime.}
    \label{fig:g2r_conv}
\end{figure}

\subsection{Envelope Mass Fraction to Planetary Radius}
\label{ssec:gcr2rad}
To compute the radius of the modeled (proto-)planets as they form, we use the interior structure models of \cite{Thorngren2016}.  These solve the equations of hydrostatic equilibrium, mass conservation, and the equation of state (EOS) on a 1-dimensional grid.  For the EOS, we use \cite{Chabrier2019} for the H/He envelope, \cite{Thompson1990} for the rock portion of the core, and \cite{Lyon1992} for the iron portion; we assume an Earth-like 2-to-1 ratio of rock to iron in the core.  To determine the specific entropy of the envelope, we use the \cite{Fortney2007} atmosphere models to estimate the heat escaping the planet interior through the atmosphere.  In the case of extremely thin (and therefore optically thin) envelopes, this underestimates the cooling rate of the planet; fortunately, these objects' radii are dominated by the core, which is insensitive to temperature.  The core is assumed to be isothermal, with the specific heat set to $7.5 \times 10^6$ erg g$^{-1}$ K$^{-1}$.  For fast computation, we assembled a grid of evolution model radii at 40 core masses between 0.3 and 60 $M_\oplus$, 33 gas-to-core mass ratios between $10^{-6}$ and $1000$, 9 incident fluxes from $0.01 F_\oplus$ to $100 F_\oplus$ where $F_\oplus$ is the flux received by Earth from the Sun, and 100 ages from 10 Myr to 10 Gyr.

There are parts of the grid where the radius calculation fails due to the lack of coverage by the H/He EOS (mainly low mass cores with thin envelopes at late times). A simple extrapolation of the grid towards $M_{\rm gas}/M_{\rm core} \ll 10^{-3}$ results in enlarged planetary radius even at these negligible envelope mass fraction. In order to smoothly connect the grid calculation to the radius of the bare core at very low envelope mass fractions, we stitch the analytic method outlined in \citet{Lee21}, their Section 2.3, at $M_{\rm gas}/M_{\rm core} \leq 10^{-3}$, normalized to match the rest of the grid calculation. 

In our analytic calculation of planetary radii, the envelope is assumed to have shrunk to a shell of convective gas with adiabatic index of $7/5$ with a thin radiative atmosphere on top.\footnote{We note that the adiabatic index of the inner envelope of our planets when they are accreting gas is different (and computed self-consistently using realistic equation of state) from 7/5 assumed here where planets are expected to have cooled down over $\sim$0.1--few Gyrs.} In order to make a correction for this atmosphere, we add to the radius of the rcb $R_{\rm rcb}$ the photospheric radius,
\begin{equation}
    R_{\rm phot} = \ln\left(\frac{\rho_{\rm rcb}}{\rho_{\rm ph}}\right)\frac{kT_{\rm eq}}{\mu_{\rm atm}m_H g}
\end{equation}
where $\rho_{\rm rcb}$ is density at the rcb, $\rho_{\rm ph}=(2/3)\mu_{\rm atm}m_H g/kT_{\rm eq}\kappa$ is the density at the photosphere, $\mu_{\rm atm}=2.374$ the atmospheric mean molecular weight, $T_{\rm eq}$ the equilibrium temperature, $g=GM_{\rm core}/R_{\rm rcb}^2$ the surface gravity, and $\kappa=10^C\rho^{\alpha}_{\rm rcb}(k/\mu_{\rm atm}m_H)^\alpha T_{\rm eq}^{\alpha+\beta}$ is the opacity. We take $C=-7.32, \alpha=0.68, \beta=0.45$ from \citet{rogers10-general}, which is appropriate for dust-free envelopes. Following the method of \citet{Lee21} obtains unphysically high $\rho_{\rm rcb} > 1\,{\rm g\,cm^{-3}}$ for very thin envelopes (e.g., thinner than Earth atmosphere), which is a byproduct of enforcing a simple adiabatic structure. To avoid this, we take the minimum between the calculated $\rho_{\rm rcb}$ and the approximate calculation of the gas density at the bottom of the envelope:
\begin{equation}
    \rho_{\rm bottom} \sim \frac{M_{\rm gas}}{4\pi R_{\rm rcb}^2} \frac{g \mu_{\rm atm} m_H}{k T_{\rm eq}}.
\end{equation}
We show in Figure \ref{fig:g2r_conv} the result of this stitching as well as the comparison of the grid calculation presented here against the analytic calculation of \citet{Lee21} and the extrapolation of grid from \citet{Lopez14}.

\begin{figure}
    \centering
    \includegraphics[width=0.5\textwidth]{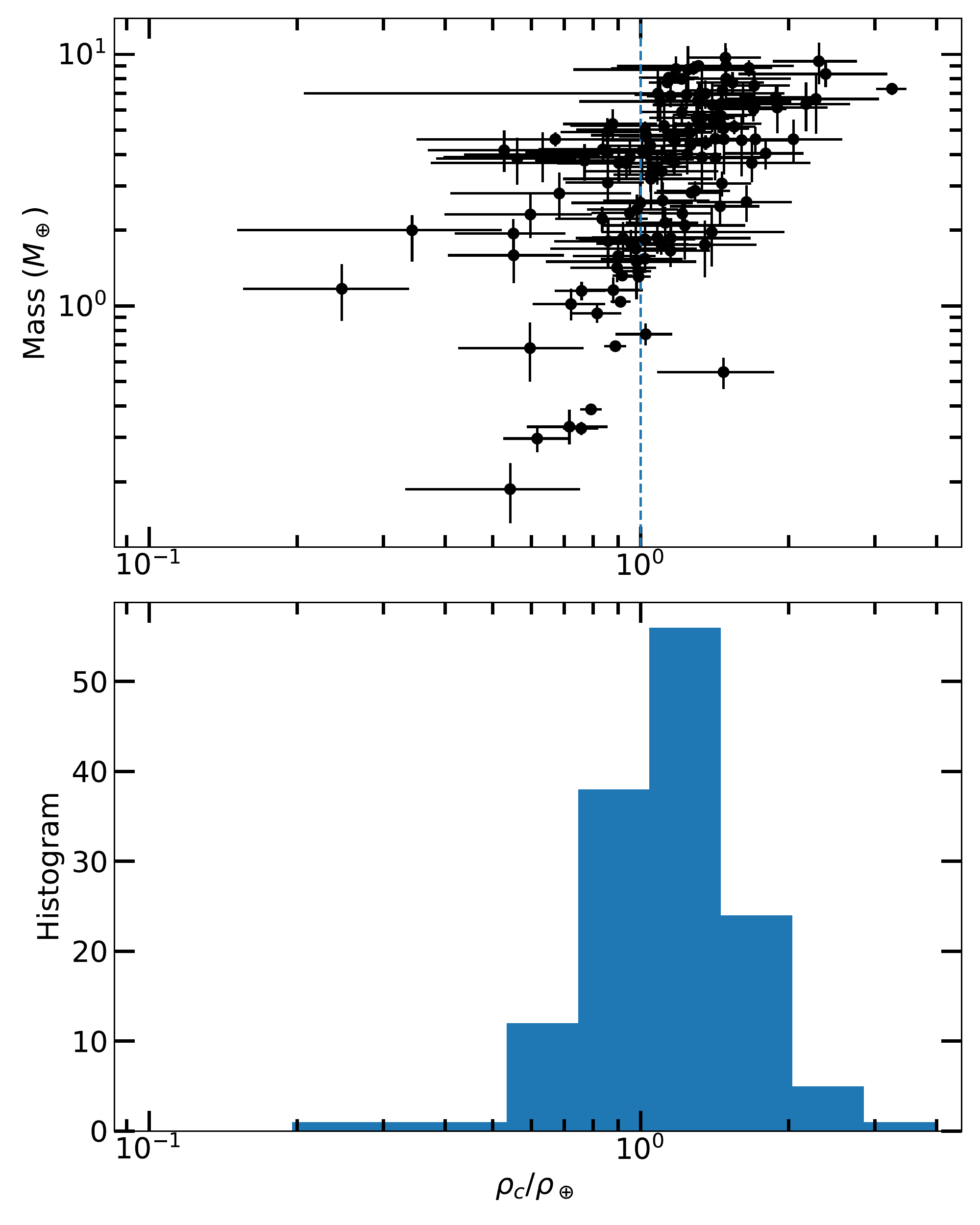}
    \caption{Distribution of small planet densities computed using the masses and radii drawn from \citet{ps}, downloaded August 21 2022. We filter for planets smaller than 2$R_\oplus$ and 10$M_\oplus$ and for those with mass and radius error estimates that are within 30\%.}
    \label{fig:rhoc_distrb}
\end{figure}

\begin{figure*}
    \centering
    \includegraphics[width=\textwidth]{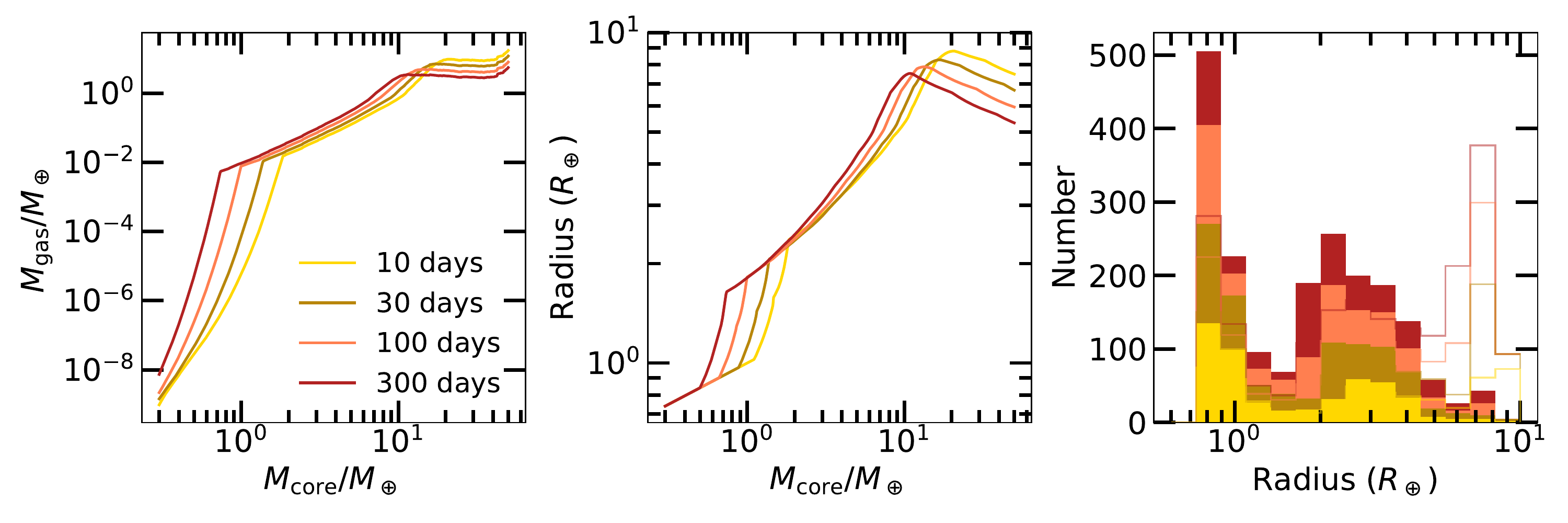}
    \caption{Left: final gas mass for each core mass. We see the steep rise in $M_{\rm gas}$ for $M_{\rm core} \lesssim M_\oplus$ from the isothermal limit and a slow rise in $M_{\rm gas}$ up to $M_{\rm core} \sim 10M_\oplus$ from accretion by cooling. After a brief sharp rise from runaway cooling, we see a plateau in $M_{\rm gas}$ from global disk accretion. The uptick in $M_{\rm gas}$ at $\sim$50$M_\oplus$ arises from the gas accretion onto these massive cores being always limited by global disk accretion whose rate is significantly higher at early times. Middle: corresponding radius (at 3 Gyr) of the $M_{\rm gas}$ curves shown on the left. Right: histogram of radius for log-uniform $M_{\rm core}$ distribution (unfilled histogram) and for cores drawn from equation \ref{eq:dNdlogMc} with $\delta = 0.3$ and $M_{\rm break}=4M_\oplus$ (filled histogram).}
    \label{fig:gcr_v_Mc_Rhist_Per}
\end{figure*}

\begin{figure*}
    \centering
    \includegraphics[width=\textwidth]{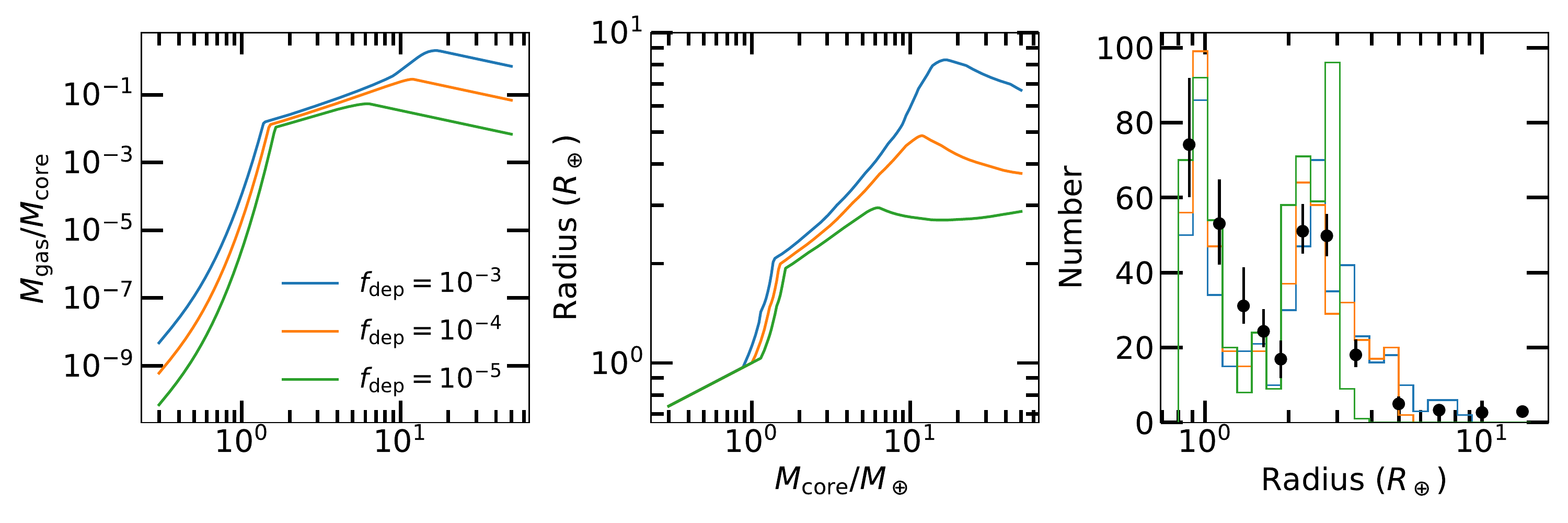}
    \caption{Same as Figure \ref{fig:gcr_v_Mc_Rhist_Per} at orbital periods of 30 days for different level of disk gas depletion at the time of gas accretion. On the right panel, we plot the histogram of the radius of 1000 model planets from 5 to 64 days with the data from \citet{Hsu19} (same range of orbital periods) plotted in black circles, rescaled for better comparison with the models. The chosen range of orbital periods maximizes the number of data points that can be plotted (i.e., actual measurements of occurrence rate rather than upper limits).}
    \label{fig:gcr_Rp_Mc_coredep}
\end{figure*}

\subsection{Core Density}
\label{ssec:core_rho}
While our radius calculation is anchored to the core composition that simulates that of the Earth, rocky exoplanets have been observed to feature a variety of core composition and therefore densities (e.g., \citealt{Zeng19}; see also \citealt{Rodriguez-Martinez22} their Figure 5). To estimate the distribution of expected exoplanet core densities, we use the data from NASA Exoplanet Archive and filter for planets smaller than 2$R_\oplus$ and 10$M_\oplus$ and for those with mass and radius error estimates that are within 30\%. Figure \ref{fig:rhoc_distrb} draws the calculated densities of these small planets $\rho_c$ in units of Earth density. We find a median $\log(\rho_c/\rho_\oplus) = 0.141$ and logarithmic standard deviation of 0.361 and so we draw the core densities from a lognormal distribution of the same median and standard deviation. In correcting our radius calculation, we simply subtract $M_c^{1/4}$ from the radius grid calculated as described in Section \ref{ssec:gcr2rad} and add $M_c^{1/4} \rho_c^{-1/3}$ for each planet. We verify that the detailed equation of state modeling does produce the core radius that scales as $M_c^{1/4}$ over the range of core masses we study.

\section{Results}
\label{sec:results}

\subsection{Radius Gap and Peak}

Figure \ref{fig:gcr_v_Mc_Rhist_Per} illustrates how gas accretion processes can naturally give rise to gaps and peaks in the distribution of planetary radii. In terms of $M_{\rm gas}$ vs.~$M_{\rm core}$ (left panel), we observe five distinct tracks. Below $\sim$1$M_\oplus$, $M_{\rm gas}$ is limited by the maximal isothermal cooling limit which is exponentially steep with $M_{\rm core}$. Between $\sim$1 and $\sim$10$M_\oplus$, $M_{\rm gas}$-$M_{\rm core}$ is shallower following the accretion by cooling relation. 
These two tracks agree with Figure 3 of \citet{Lee21}. Compared to their fiducial model, our isothermal to cooling transition appears at slightly smaller core masses because our disk temperature is colder. We also extend our calculation to larger core masses to probe the entire radius distribution beyond $\sim$3--4$R_\oplus$. 
We see a short hint of runaway accretion at $M_{\rm core} \sim 10 M_\oplus$ beyond which we see a plateau in $M_{\rm gas}$ set by the global disk accretion, which is independent of core mass. There is a short uptick at $\sim$50$M_\oplus$ caused by the gas accretion onto these massive cores being always limited by global disk accretion and the rate of disk accretion is higher at earlier times. 

The $M_{\rm gas}$-$M_{\rm core}$ tracks are closely followed by the radius-$M_{\rm core}$ relationship (middle panel) with distinct differences at the lowest and at the highest end of $M_{\rm core}$. Once $M_{\rm gas}$ falls below $\sim$10$^{-4} M_\oplus$, the radius is given by that of the core at all orbital periods. At the high mass end when the final $M_{\rm gas}$ is set by the disk accretion and therefore is the same for all core masses, we see a downturn in radius with core mass because the envelope mass {\it fraction} decreases with core mass. The total accreted gas mass in this regime is larger at shorter orbital periods because the onset of gas accretion starts earlier (see equation \ref{eq:tstart}) and so the rate of cooling exceeds the global disk accretion rate at earlier times when the latter is higher. 

\begin{figure}
    \centering
    \includegraphics[width=0.5\textwidth]{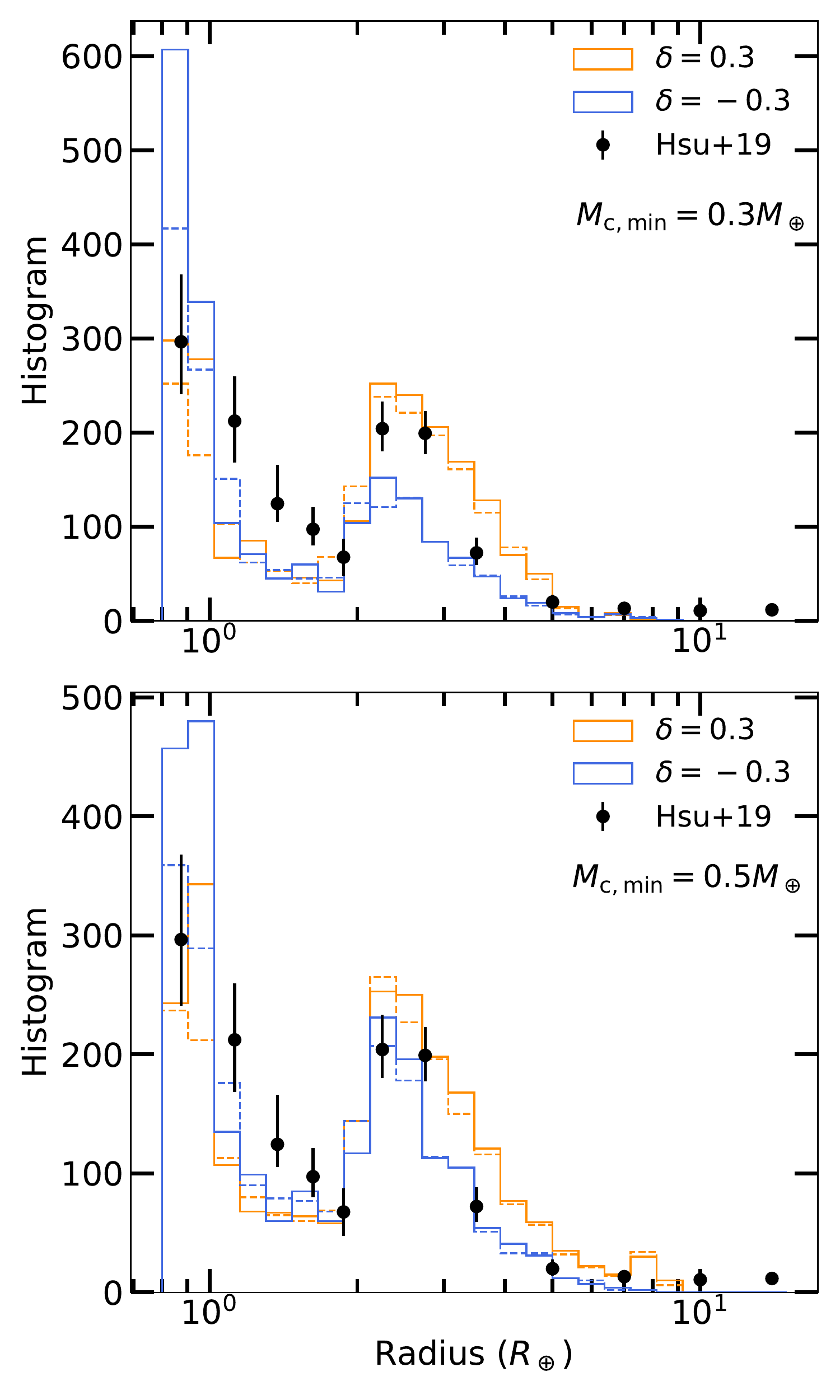}
    \caption{Radius distribution of 5000 planets (evaluated at 3 Gyrs) for bottom-heavy ($\delta=-0.3$) and top-heavy ($\delta=0.3$) core mass distributions with $M_{\rm break}=4M_\oplus$ with a cut-off in the low mass end $M_{\rm c,min}$ at 0.3$M_\oplus$ (top) and 0.5$M_\oplus$ (bottom). While both distributions consistently produce a gap at $\sim$1--2$R_\oplus$ and a peak at $\sim$2--3$R_\oplus$, the radius peak of the $\delta=-0.3$ is sharper and is shifted slightly towards lower radius. Plotted in dashed lines are the correction to the radius distribution accounting for the variance in core density (see text for more detail). We have rescaled the data points of \citet{Hsu19} for better comparison.}
    \label{fig:Rd_Mcfunc}
\end{figure}

These sudden rises and lulls in the radius-mass scaling relationship translate directly into gaps and peaks in the radius distribution. In the right panel of Figure \ref{fig:gcr_v_Mc_Rhist_Per}, we see a sharp gap at $\sim$1--2$R_\oplus$ due to the exponential rise of $M_{\rm gas}$ and radius with $M_{\rm core}$ in the isothermal limit (i.e., there is not enough dynamic range in $M_{\rm core}$ to populate 1--2$R_\oplus$). The downturn in radius at $M_{\rm core} \gtrsim 10M_\oplus$ gives rise to a radius peak at $\sim$10$R_\oplus$ for a uniform core mass distribution. As we established in Section \ref{ssec:mass_time_per}, planetary core mass function is more likely skewed towards low mass with a few cores larger than 10$M_\oplus$. Drawing $M_{\rm core}$ from equation \ref{eq:dNdlogMc} using $\delta=0.3$ and $M_{\rm break}=4M_\oplus$ (filled histogram) we find the radius peak at $\sim$2--3$R_\oplus$, in agreement with the observed exoplanet radii distribution. In this picture, the fall-off in the number of Sub-Saturns and gas giants is a direct reflection of the sharp fall-off in the underlying core mass distribution at the high mass end. 

Drawing the planetary orbital periods from equation \ref{eq:dNdlogP}, we find that over the range of disk gas depletion factor that can trigger core assembly by mergers, the gap in the radius distribution at $\sim$1--2$R_\oplus$ and the peak at $\sim$2--3$R_\oplus$ appear consistently, as demonstrated in Figure \ref{fig:gcr_Rp_Mc_coredep}. At the strongest depletion $f_{\rm dep}=10^{-5}$, there is a pronounced sharp peak close to $\sim$3$R_\oplus$ generated by planets with core masses $\gtrsim$5$M_\oplus$ that attain their final gas mass fraction of a few to ten percent by reduced global disk accretion (see left and middle panels of Figure \ref{fig:gcr_Rp_Mc_coredep}). The fact that this peak is not seen in the data suggests that, in the absence of post-formation processes further sculpting the radius distribution, most cores complete their assembly process when the disk gas depletion level reaches $10^{-4}$--$10^{-3}$. We will show later in Section \ref{ssec:mloss} that if mass loss processes are active, then a population of planets created at $f_{\rm dep}=10^{-5}$ is required. In the sections where we describe the main result of primordial gas accretion in shaping the exoplanetary population, we will focus on $f_{\rm dep}=10^{-3}$ and return to other $f_{\rm dep}$ in Section \ref{sec:disc}.

We also illustrate in Figure \ref{fig:Rd_Mcfunc} how the gap and peak in the radius distribution appear stable between more top-heavy ($\delta=0.3$) and bottom-heavy ($\delta=-0.3$) core mass distributions. For $\delta=-0.3$, the radius peak is sharper and shifts slightly towards smaller radii ($\sim$2$R_\oplus$); we also see a stronger concentration of sub-Earth size objects as expected for a bottom-heavy core mass distribution. While such a strong concentration is not seen in the observations, we find that correcting the expected radii of model planets for a variety of possible core densities (Section \ref{ssec:core_rho}) can smear out sub-Earth concentration to a better agreement to that observed (see dashed histograms in Figure \ref{fig:Rd_Mcfunc}). For the bottom-heavy core distribution, the relative size of the radius peak at $\sim$2--3$R_\oplus$ shrinks when we cut off our core mass distribution at the lowest mass end at 0.3$M_\oplus$ instead of 0.5$M_\oplus$ since this implies a larger population of sub-Earth objects. By contrast, $\delta=0.3$ runs are insensitive to where we set the minimum core mass since these small planet population were a relative minority to begin with. A better observational constraint on the shape of the radius distribution below an Earth radii would help determine whether such a cut-off in the low mass end is required or not. For this paper, we adopt $M_{\rm c,min}=0.5M_\oplus$ as our fiducial parameter and correct the radius calculations for a variety of core densities. 

\begin{figure}
    \centering
    \includegraphics[width=0.5\textwidth]{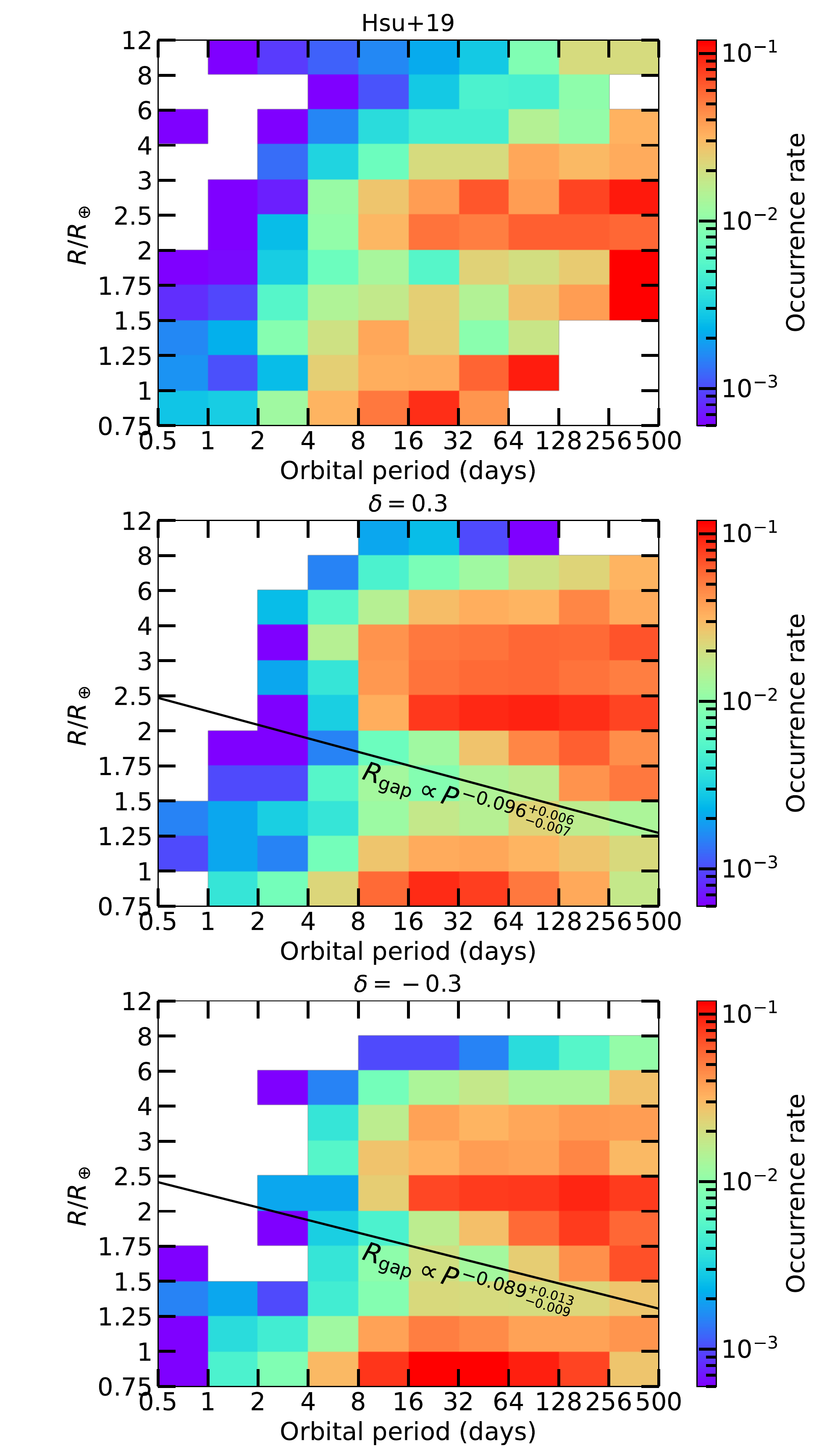}
    \caption{Radius-period distribution of \citet{Hsu19} data (top) and from the accretion model (bottom). Overplotted in the bottom two panels are the best power-law fits to the shape of the radius gap.}
    \label{fig:RvP_obs_prim}
\end{figure}

\subsection{Radius-Period Distribution}
\label{ssec:RP}

As illustrated in Figure \ref{fig:RvP_obs_prim}, our fiducial accretion models reproduce the location and the overall morphology of the observed radius-period distribution of {\it Kepler} planets, including the shape of the ``gap'' that separates planets smaller than $\sim$1.5$R_\oplus$ and the fall-off in the number of planets larger than 4$R_\oplus$ at all orbital periods. 

Using the \texttt{gapfit} routine of \citet{Loyd20}, we obtain $R_{\rm gap} \propto P^{-0.096^{+0.006}_{-0.007}}$ for $\delta=0.3$ and $R_{\rm gap} \propto P^{-0.089^{+0.013}_{-0.009}}$ for $\delta=-0.3$. Both fits are well within 1-$\sigma$ agreement with $R_{\rm gap} \propto P^{-0.09^{+0.02}_{-0.04}}$ reported by \citet{vanEylen18} and with $R_{\rm gap} \propto P^{-0.11\pm0.02}$ reported by \citet{Martinez19}. As described in \citet{Lee21} the slightly decreasing $R_{\rm gap}$ with orbital period can be understood by the enlargement of the maximal isothermal mass limit for a given core mass farther from the star where the disk is colder. The primordial radius gap arises at the transition core mass between the isothermally-limited to envelope growth by cooling and so larger isothermal mass results in smaller transition core mass and therefore smaller $R_{\rm gap}$ (see also Figure \ref{fig:gcr_v_Mc_Rhist_Per}). The drop in the number of planets with radii 0.75--1$R_\oplus$ beyond $\sim$128--256 days we see in our model population reflects this drop in transition core mass at longer orbital periods. At such long orbital periods, the disk temperature is low enough for even the small cores ($\lesssim M_\oplus$) to accrete sufficient amount of gas to have their radii blow up approaching $\sim$2$R_\oplus$.

Both in the data (top panel of Figure \ref{fig:RvP_obs_prim}) and in the model (the bottom two panels of Figure \ref{fig:RvP_obs_prim}), we find an upper envelope in the radius-period distribution where we find more planets with radii $\gtrsim$2.5$R_\oplus$ at longer orbital periods. This envelope corresponds to the lower end of the sub-Jovian desert \citep[e.g.,][]{Mazeh16} which can be sculpted by photoevaporation \citep[e.g.,][]{Lundkvist16,Owen18,Hallatt22}. We find that such an envelope also appears purely from gas accretion for an underlying core mass distribution that falls off beyond a few $M_\oplus$. Under dust-free accretion, the opacity drops at longer orbital periods and so gas accretion accelerates; runaway is triggered for lower mass cores; and the disk-limited accretion applies to a wider range of core masses (see the left panel of Figure \ref{fig:gcr_v_Mc_Rhist_Per}), naturally allowing for larger population of Neptune to Saturn-sized objects at longer orbital periods. We note that the same behavior is expected even in dusty accretion as long as the opacity drops at longer orbital periods, as was found by \citet{Chachan21} who computed in detail the effect of dust dynamics on the overall opacity gradient in protoplanetary disks.

\begin{figure}
    \centering
    \includegraphics[width=0.5\textwidth]{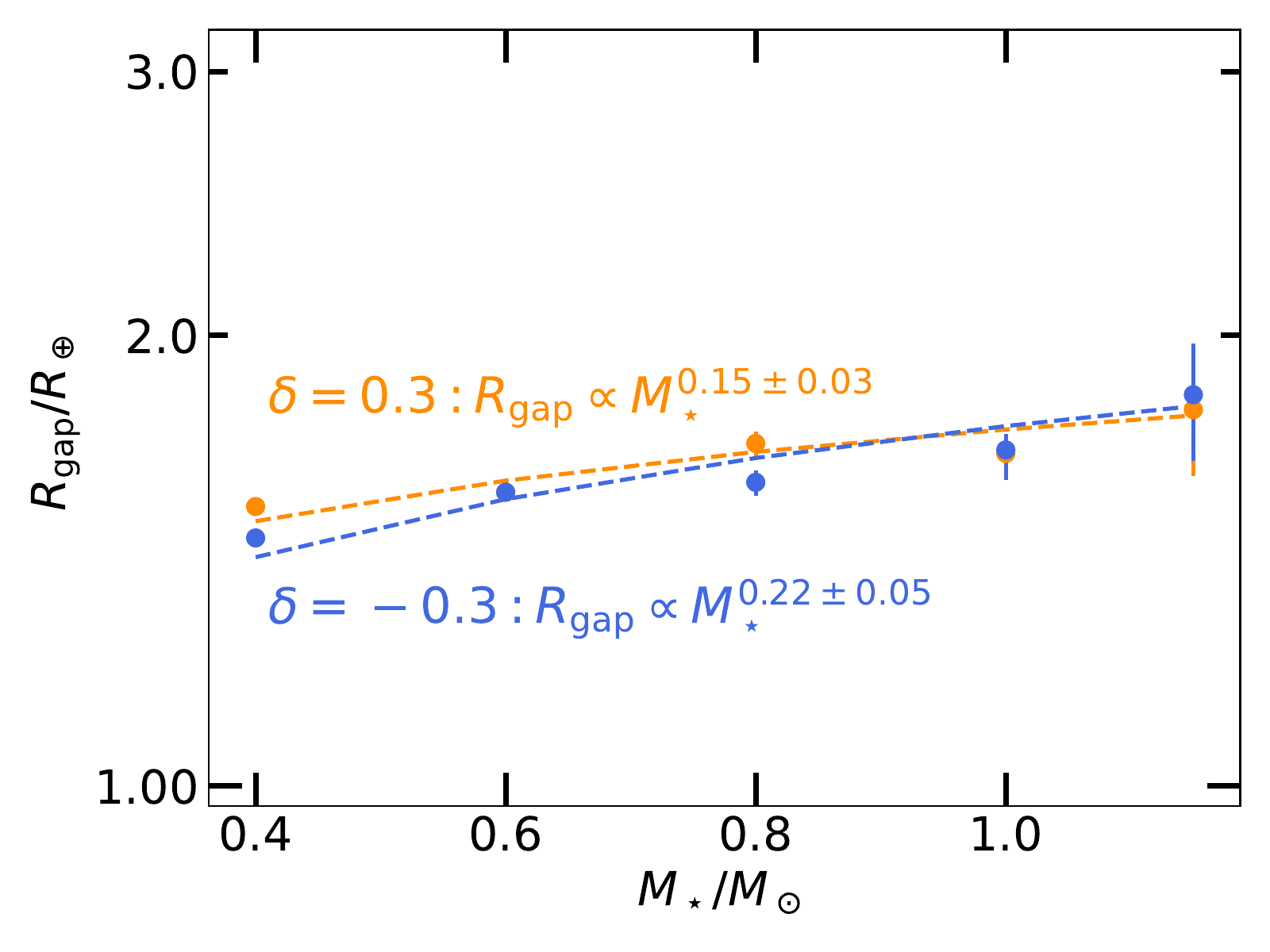}
    \caption{The location of radius gap evaluated at an orbital period of 30 days for each stellar mass. In general, the gap shifts to larger radii around more massive stars for both bottom-heavy ($\delta=-0.3$) and top-heavy ($\delta=0.3$) distributions. The best-fit scaling relationship between the gap radius $R_{\rm gap}$ and the stellar mass $M_\star$ for both populations agree with the measurement by \citet{Berger20} ($R_{\rm gap}\propto M_\star^{0.26^{+0.21}_{-0.16}}$) and \citet{Petigura22} ($R_{\rm gap}\propto M_\star^{0.18^{+0.08}_{-0.07}}$) within 1-$\sigma$.}
    \label{fig:Rgap_v_Mstar}
\end{figure}

\subsection{Radius-Stellar Mass}

Around hotter (massive) stars, the disks are hotter, so we expect the radius gap that emerges from primordial gas accretion to shift to larger radius around hotter stars. From simple scaling relations, \citet{Lee21} deduced $R_{\rm gap} \propto M_\star^{0.11}$ for the primordial radius gap arising in passively-heated disks. To verify our expectations, we generate model planet population around stars of varying masses drawing their appropriate bolometric luminosity evolution from the \citet{Johnstone21} grid. Furthermore, we scale $\dot{M}_{\rm disk} \propto M_\star^{1.95}$ from \citet{Calvet04} and $\Sigma_{\rm bg} \propto M_\star$ from \citet{Andrews13} (assuming the disk solid and gas mass scale the same way with $M_\star$). Since $\dot{M}_{\rm disk} \propto \nu \Sigma_{\rm bg}$, the kinematic viscosity $\nu \propto M_\star^{0.95}$ and so $t_{\rm visc} \propto M_\star^{-0.95}$ which implies the disks around lower mass stars are expected to live longer but that the gas accretion onto planetary cores also begins later.

Our expectation is corroborated in Figure \ref{fig:Rgap_v_Mstar} where we quantitatively solve for the location of the radius gap in radius-period space (following the procedure outlined in Section \ref{ssec:RP}) for model planet population around each stellar mass. A least-square fit finds that the shift in gap radius with stellar mass is modest with $R_{\rm gap} \propto M_\star^{0.15\pm0.03}$ for the top-heavy ($\delta=0.3$) fiducial model and $R_{\rm gap} \propto M_\star^{0.22\pm0.05}$ for bottom-heavy ($\delta=-0.1$) fiducial model, both of which are in 1-$\sigma$ agreement with the measurement by \citet{Berger20} who quote $R_{\rm gap} \propto M_\star^{0.26^{+0.21}_{-0.16}}$ and by \citet{Petigura22} who find a shallower dependence $R_{\rm gap} \propto M_\star^{0.18^{+0.08}_{-0.07}}$.

\begin{figure}
    \centering
    \includegraphics[width=0.5\textwidth]{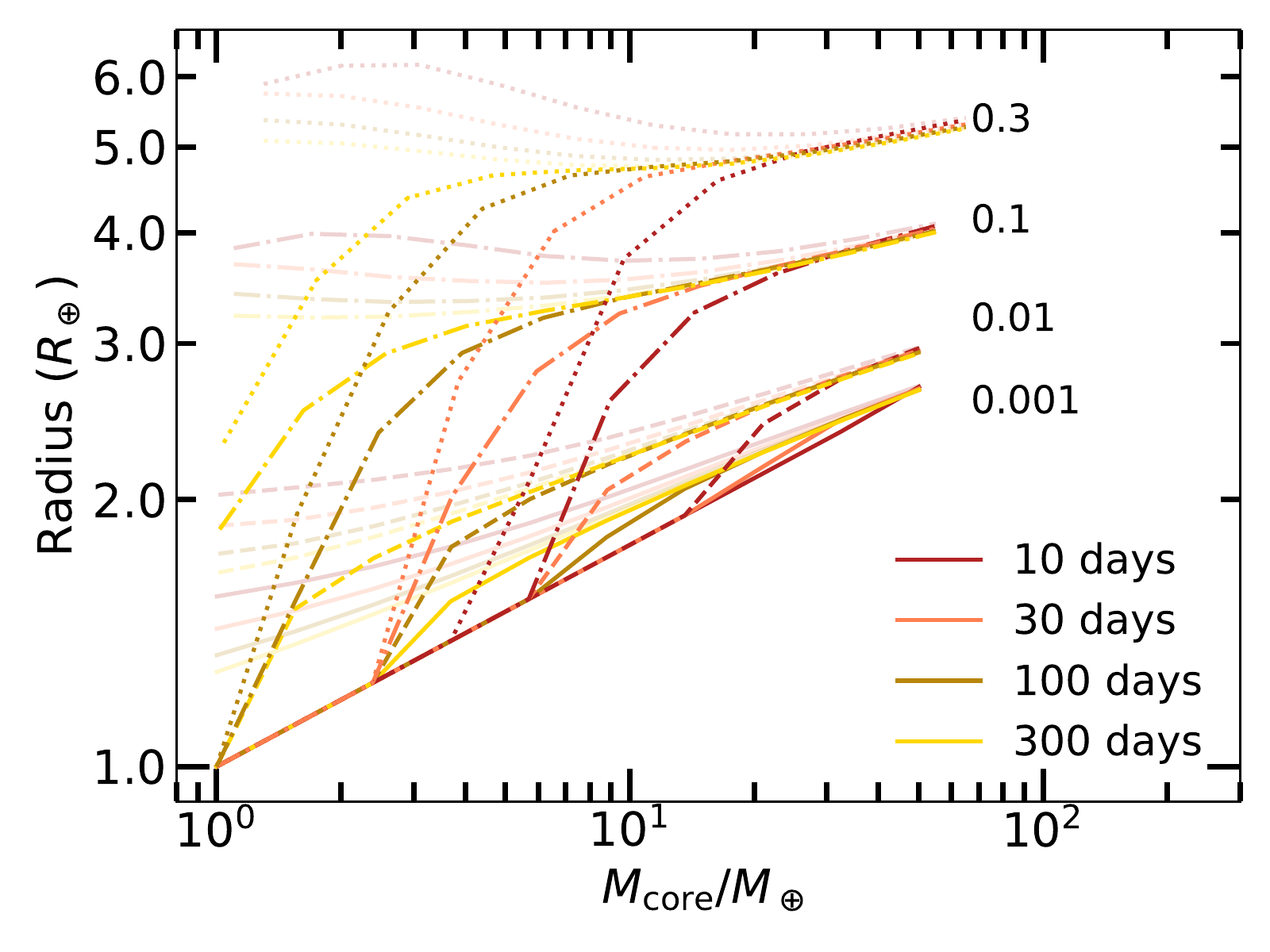}
    \caption{Radius vs.~core mass for varying initial envelope mass fraction $M_{\rm gas,init}/M_{\rm core}$ and orbital periods. Plotted in dark lines are the radii after photoevaporation over 3 Gyr whereas plotted in light lines are the radii from pure thermal evolution over 3 Gyr without mass loss. To the right of each family of lines (different line styles), we indicate their corresponding $M_{\rm gas,init}/M_{\rm core}$.}
    \label{fig:RvM_evap_diag}
\end{figure}

\begin{figure*}
    \centering
    \includegraphics[width=\textwidth]{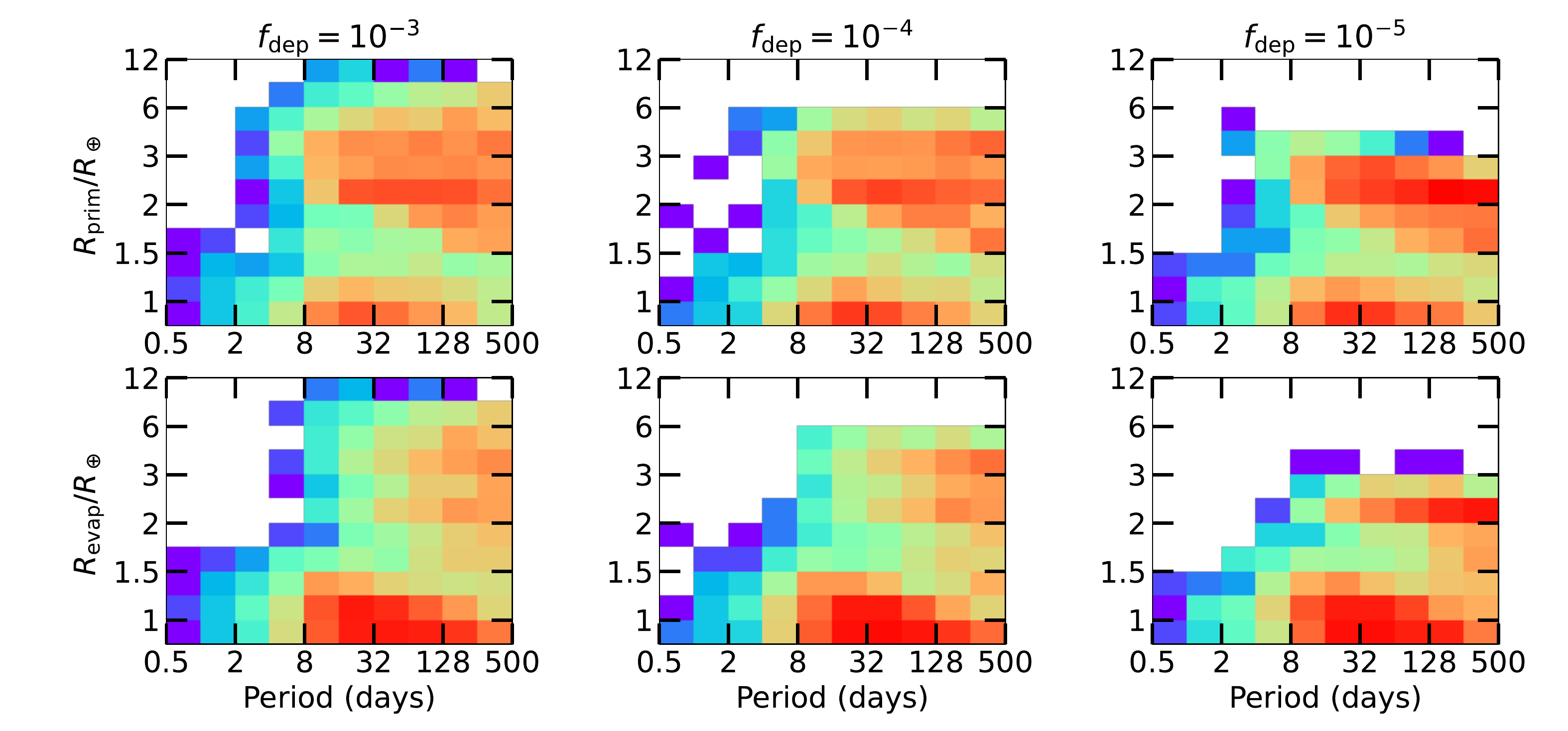}
    \caption{Radius-period distribution for primordial population thermally evolved to 3 Gyr (top) and for those undergone photoevaporative mass loss for 3 Gyr (bottom). The underyling core mass distribution follows $\delta = 0.3$ and $M_{\rm break}=4M_\oplus$.}
    \label{fig:RvP_prim_evap}
\end{figure*}

\section{Discussion}
\label{sec:disc}

We have demonstrated that with more careful accounting of the disk temperature profile (which begets a colder disk), the primordial radius gap hypothesized by \citet{Lee21} remains a viable explanation for the observed gap seen in the radius distribution and in the radius-period space, even with an underlying core mass distribution that is extended to sub-Earth mass regime and sometimes bottom-heavy ($\delta=-0.3$). The ability to create the radius gap even with bottom-heavy core mass distribution updates the finding of \citet{Lee21} who concluded that the underlying core mass distribution can be broad but not broader than $dN/dM_{\rm core} \propto M_{\rm core}^{-0.7}$ equivalent to our $\delta=0.3$. This difference mainly arises from our choice of reference observations by \citet{Hsu19} instead of \citet{Fulton18} and also our choice of using the raw histogram instead of smoothing by Gaussian kernel which is known to smooth out fine signatures like the observed radius gap in 1-dimensional radius space. 

In this section, we discuss the effect of post-formation processes on further shaping the exoplanetary population and discuss avenues for future work.

\subsection{Effect of Mass Loss}
\label{ssec:mloss}
The observed shift in the radius distribution of exoplanets across $\sim$1--2 Gyr suggests that there must be a post-formation process that operates over Gyrs' timescale sculpting the exoplanetary population \citep[e.g.,][]{Berger20,David21}. Although initially, photoevaporation was considered to be only effective over $\sim$100 Myrs, recent updates to our understanding of EUV spectrum of stars suggest this channel of mass loss may well be active over a Gyr timescale. Core-powered envelope mass loss can also naturally give rise to mass loss over Gyrs' timescale. However, the ability of the cores to cool down fast enough has recently been called into question \citep[e.g.,][]{Markham22}, and so we focus on the effect of photoevaporation.

We use the empirical formula fitted to detailed hydrodynamic calculations provided by \citet{Caldiroli21}, whose expression encompasses both the energy-limited regime and the non-energy-limited regime:
\begin{equation}
    \dot{M} = -\eta \frac{\pi R_p^3 F_{\rm XUV}}{G K M_p}
\end{equation}
where $\eta$ is the efficiency factor that varies non-trivially with the irradiation flux and the planet's gravitational potential (see \citealt{Caldiroli21}, their Appendix A.1), $R_p$ is planetary radius using the conversion method we described in Section \ref{ssec:gcr2rad}, and $K= 1 - 1.5 (R_{\rm Hill}/R_p)^{-1} + 0.5 (R_{\rm Hill}/R_p)^{-3}$ is the reduction factor to the potential due to tidal effects from the star \citep{Erkaev07}. It is the complex behavior of $\eta$ that ensures the prescription we use accounts for the non-energy-limited regime. To calculate the XUV flux $F_{\rm XUV}$, we use the empirical relation of X-ray vs.~time and the X-ray to EUV ratio reported by \citet{King21}. 

The X-ray luminosity is given by
\begin{equation}
    L_{X} = 
    \begin{cases}
    L_{\rm sat} & t < 100\,{\rm Myrs}\\
    L_{\rm sat}\left(\frac{t}{100\,{\rm Myrs}}\right)^{-1.4} & t \geq 100\,{\rm Myrs}
    \end{cases}
\end{equation}
where $L_{\rm sat} = 10^{-3.6}L_\star(t)$ is the saturation luminosity and $L_\star(t)$ is the time-dependent bolometric luminosity taken from the stellar grid of \citet{Johnstone21}.

The EUV luminosity scaling differs between hard (100--360 {\AA}) and soft (360--920 {\AA}) bands. Following \citet{King21}, we adopt
\begin{equation}
    L_{\rm EUV,hard} = 116\left(\frac{L_\star(t)}{4\pi R^2_\star(t)}\right)^{-0.35}\left(\frac{L_{X}(t)}{L_\star(t)}\right)^{0.65}
\end{equation}
for hard EUV and 
\begin{equation}
    L_{\rm EUV,soft} = 3040\left(\frac{L_\star(t)}{4\pi R^2_\star(t)}\right)^{-0.76}\left(\frac{L_{X}(t)}{L_\star(t)}\right)^{0.24}
\end{equation}
for soft EUV, with all quantities evaluated in cgs unit. 

As illustrated in Figure \ref{fig:RvM_evap_diag}, mass loss over 3 Gyr expects cores of masses $\sim$5$M_\oplus$ with initial envelope mass fraction $\gtrsim$0.1 to appear as sub-Neptunes ($\sim$2--3$R_\oplus$) over orbital periods of $\sim$10--30 days. The required initial envelope mass fraction is about an order of magnitude higher than what was reported by \citet{Owen17} because $\eta$ is often a few times higher than their assumed 0.1 \citep[see also][]{Kubyshkina18} and our updated XUV luminosity does not fall off as steeply as assumed in their work. 

Figure \ref{fig:RvP_prim_evap} shows the effect of mass loss over 3 Gyr on the radius-period distribution for cores that have assembled at different disk gas depletion factors (equivalently, at different times in disk evolution). Mass loss consistently removes $\gtrsim$2--4$R_\oplus$ planets into $\lesssim$1.5$R_\oplus$ planets, filling up the radius gap on the edge of small radius ($\sim$1.5$R_\oplus$) and carving it out on the edge of large radius ($\sim$2$R_\oplus$). By comparison to the occurrence rate calculations by \citet{Hsu19} on the radius-period plane (top panel of Figure \ref{fig:RvP_obs_prim}), we deduce that the shape of the radius peak at $\sim$2--3$R_\oplus$ under photoevaporative mass loss is best explained if most cores started accreting gas at $f_{\rm dep}=10^{-5}$ and the population of large planets ($\gtrsim$4$R_\oplus$) was created earlier at $f_{\rm dep}=10^{-3}$.

\begin{figure}
    \centering
    \includegraphics[width=0.5\textwidth]{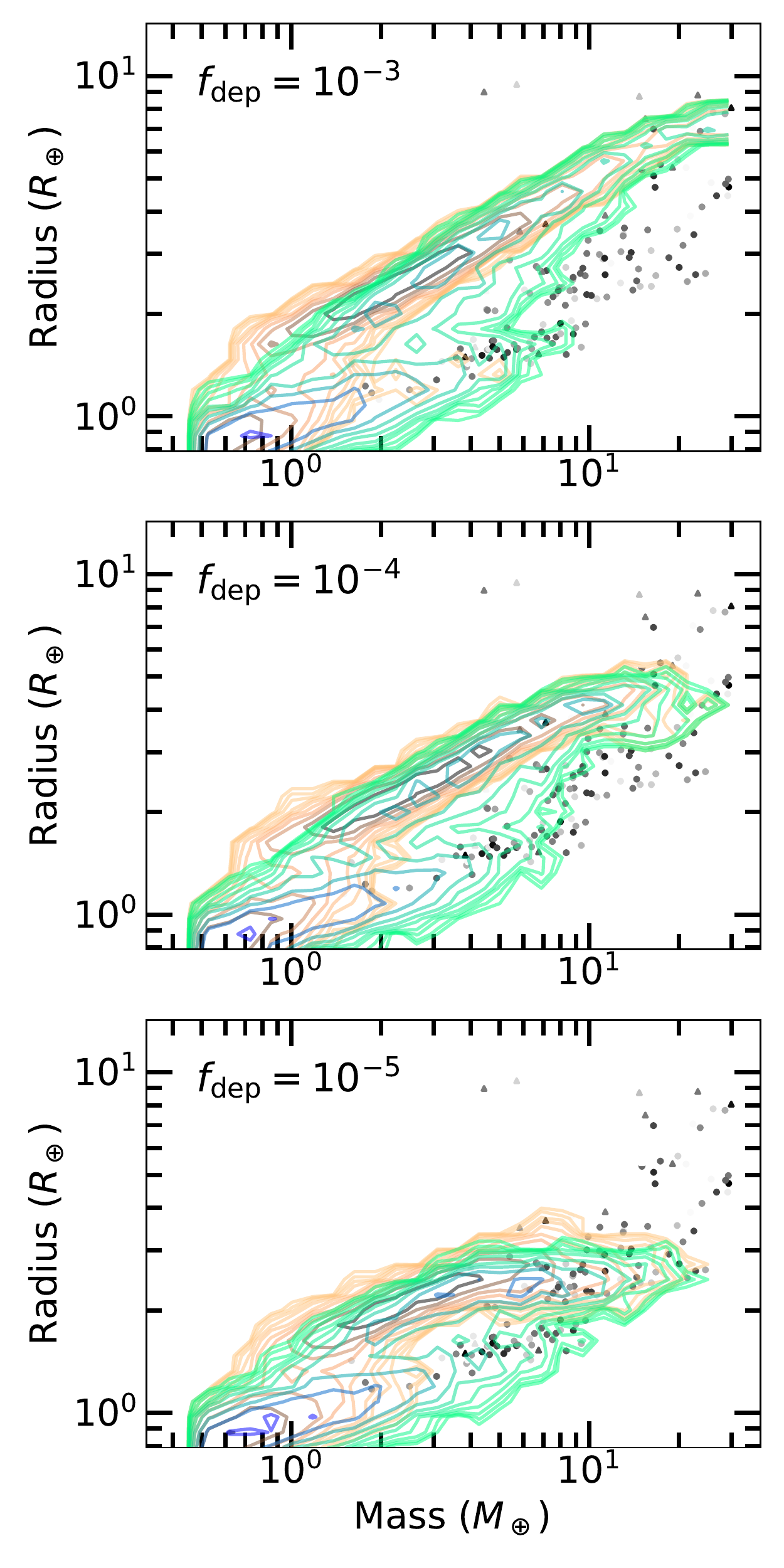}
    \caption{Radius-total mass distribution in contour lines for the primordial population (orange) and for the photoevaporated population (green) for different $f_{\rm dep}$, both evolved to 3 Gyr. We plot the data from \citet{ps21} for mass measurement from radial velocity (circles) and from transit timing variation (triangles) with lighter colors denoting those with larger percentage error. The datapoints are filtered to those with measurement errors within 30\%, orbital periods less than 500 days, masses less than 30$M_\oplus$ and radius less than 10$R_\oplus$. The two triangles of mass $\sim$4--6$M_\oplus$ and radius $\sim$10$R_\oplus$ are Kepler-51c,d which are known super-puffs that require special formation channels including gas accretion in beyond $\sim$1 AU \citep[e.g.,][]{Lee16}.}
    \label{fig:RvM_prim_evap}
\end{figure}

\subsection{Mass-Radius Distribution}

In mass-radius space, primordial gas accretion alone produces two tracks (see the orange contour lines in Figure \ref{fig:RvM_prim_evap}): the lower track corresponding to bare cores $R \propto M^{1/4}$ and the upper track corresponding to gas-enveloped planets with the space between the two tracks ($\sim$1--2$R_\oplus$) sparsely populated, in line with the observed gap in the radius distribution. With more severe disk gas depletion, the upper track starts to veer off at larger radii around mass of $\sim$10$M_\oplus$. At $f_{\rm dep}=10^{-4}$, the upper track populates $\sim$3--4$R_\oplus$ at $\gtrsim$10$M_\oplus$ whereas for $f_{\rm dep}=10^{-5}$, the upper track populates $\sim$2--3$R_\oplus$. The final gas mass accreted by these massive cores is eventually limited by global disk accretion which is larger at earlier times (i.e., larger $f_{\rm dep}$) so we expect this trend of creating larger planets at larger $f_{\rm dep}$ for more massive planets.

We find that these two tracks from primordial gas accretion are generally stable against photoevaporation (compare orange and green contour lines in Figure \ref{fig:RvM_prim_evap}), with two notable differences. First, the low mass end of the upper track is carved out and move down to the lower track, shifting the peak of the upper track to larger masses beyond $\sim$3$M_\oplus$. Second, the high mass end of the lower track extends out to larger masses out to $\sim$10$M_\oplus$. From comparison to the observed data, we find that the locus of observed planets at $\sim$1--2$R_\oplus$ and $\sim$4--10$M_\oplus$ are likely photoevapoarted population whereas those with $\sim$2--10$R_\oplus$ and $\gtrsim$6$M_\oplus$ are the primordial population that are stable against mass loss.

Our model population is peaked at lower masses as compared to current observations. For example, planets with both measured radii and masses (with percentage error within 30\%) fill in the observed radius gap of $\sim$1--2$R_\oplus$. We note that the total number of confirmed planets with mass measurements is only about 10\% of confirmed planets with radii measurements. Given the observational bias against low mass objects, it is likely that with more mass measurements that can probe lighter planets, we may see real-life exoplanets filling in the low mass end of both the upper and the lower track of the mass-radius phase space predicted by our dust-free gas accretion models.

\subsection{System Metallicity}

Observations report a strong positive correlation between the occurrence rate of gas giants and stellar metallicity \citep[e.g.,][]{Fischer05} whereas the metallicity trend with sub-Neptune occurrence rate is significantly weaker \citep[e.g.,][]{Wang15,Petigura18,Wilson21}, with hotter planets more likely to appear around metal-rich stars \citep[e.g.,][]{Wilson18}.

In our theory of primordial gas accretion, envelope metallicity can affect the final gas mass and the apparent radius of the planet. With all else equal, higher metallicity effects slower cooling and therefore less accretion, leading to smaller planets. Once $Z \gtrsim 0.4$, the trend reverses and the high mean molecular weight enhances gas accretion, leading to larger planets. By comparison, atmospheres with larger metallicity will appear larger due to enhanced opacity.\footnote{Such extended atmospheres can enhance the mass loss rate. However, enhanced metallicity can also reduce the rate of mass loss through metallic line cooling which can make upper atmosphere cooler and so render them harder to leave the planet's gravitational potential (\citealt{Owen_Murray-Clay18}; see also \citealt{Mordasini20}).}

To see which effect dominates, we compare the expected change in radius from solar metallicity to 50$\times$ solar metallicity in number, corresponding to a change in mass ratio $Z=0.02$ to $Z \sim 0.4$ (equivalently, $\mu_{\rm env}=2.4$ to $\mu_{\rm env}=3.6$). From equation \ref{eq:gcr_tscl_df}, $M_{\rm gas}/M_{\rm core} \propto Z^{-0.4}\mu_{\rm env}^{2.2}$ and so $M_{\rm gas}/M_{\rm core}$ decreases by a factor of $\sim$0.76, which amounts to $\sim$3-4\% decrease in radius. The same enhancement in envelope metallicity enlarges the effective radius of the planet by $\sim$3\% as well \citep{Lopez14}, so changes in radii for 50$\times$ enhancement in metallicity compared to solar value are within the current measurement uncertainties in exoplanet radius (and in this case, the two effects would cancel each other out). Sufficiently large $Z \gtrsim 0.4$ would produce a population of larger planets; however, such high metallicity enhancement in planetary envelope is likely not the norm as it would push would-be sub-Neptunes into runaway accretion and create more gas giants \citep{Lee16}, contrary to the observed ubiquity of sub-Neptunes.

What is more likely is that all else is not equal with varying stellar metallicity; in fact, stellar metallicity may very well be decoupled from the elemental abundances of planetary envelopes \citep[see, e.g.,][for observational hints]{Petigura22}. Instead, we consider the present-day metallicity enhancement over solar value of planet-hosting stars reflecting the metal-enrichment of their protoplanetary disks where the more massive solid reservoir likely spawned systematically more massive cores, creating larger planets \citep[e.g.,][]{Dawson15}. Core growth by pebble accretion is also enhanced with higher disk metallicity \citep[e.g.,][]{Lambrechts14,Lin18} and the disks may disperse later around stars in high metallicity environment \citep[e.g.,][]{Yasui09}, allowing for the emergence of more gas-rich planets in solid-enhanced disks. Our conclusion here is in alignment with what \citet{Lee19} suggested in that planets larger than $\sim$4$R_\oplus$ can only appear in metal-rich disks that can spawn massive cores early enough whereas smaller planets can appear in any disks so long as the cores assemble in gas-depleted environments, giving rise to the weak metallicity-dependence of small planet occurrence rate. More mass measurements of the planets to search for stellar metallicity - planet mass trend would help test this hypothesis.

\subsection{Dusty vs.~Dust-free}
\label{ssec:dusty_v_df}

In this paper, we focussed our analysis on ``dust-free'' accretion whereby dust grains do not contribute to the opacity of either the envelope or the underlying disk. For instance, dust grains are expected to coagulate and rain out within planetary envelopes \citep[e.g.,][]{Mordasini14,Ormel14} and within $\sim$100 days of orbital period, solid coagulation timescales are orders of magnitude shorter than $\sim$10 Myrs so that most solids are likely collected into planets. While likely, it is not definitive that gas accretion be dust-free so in this section, we briefly comment on what we may expect when we account for dust opacity and defer a more careful study to a future project.

First, from Figure \ref{fig:optdepth_rho}, we see that a dusty disk will be optically thick unless the disk surface density falls below 6 orders of magnitude. The midplane temperature of these optically thick disks will be dominated by accretional heating reaching temperatures $\sim$2000 K in the inner disk for an accretion rate of $\sim 10^{-8}\,M_\odot\,{\rm yr}^{-1}$ \citep[see also][]{Jankovic21}. In hotter disks, the maximal isothermal limit will drop and the higher opacity from dust shrinks the accretion-by-cooling gas mass. These drops will push the transition core mass between the maximal isothermal limit and the cooling curve to a larger value, while keeping the overall gas content smaller. This would push down the expected upper track in the mass-radius plane and we would expect an underlying core mass distribution that fall off at higher masses than what we found with the dust-free accretion. 

\section{Conclusion}
\label{sec:concl}

Using more realistic disk conditions, we demonstrated that even in the absence of mass loss, the radius gap can naturally appear from primordial gas accretion alone as the physics that limits the rate of accretion onto planetary cores switches. Light cores ($M_{\rm core} \lesssim$1--2$M_\oplus$, depending on the orbital period) reach their isothermal maximally cooled limit, unable to accrue sufficient amount of gas so that they are fated to remain as rocky terrestrials. The isothermal-to-cooling transition carves out a gap at $\sim$1--2$R_\oplus$ in the radius distribution. Massive cores ($M_{\rm core} \gtrsim$10$M_\oplus$) trigger runaway accretion before being halted by global disk accretion. If the cores assemble late enough (when the disk gas is depleted by 5 orders of magnitude with respect to MMEN), then the paucity of planets beyond $\sim$3$R_\oplus$ is created by the limited amount of global disk accretion. If the cores assemble earlier, then this paucity requires an underlying core mass distribution that falls off sharply beyond a few Earth masses.

Accounting for these limits and transitions on gas mass allows for the underlying core mass to be broad towards low mass end and even bottom-heavy extending to sub-Earth masses so that both the rise in the occurrence rate of sub-Earth sized planets and the location of the radius valley \citep{Hsu19} can be reproduced. Our model further reproduces the shape of the radius valley in the radius-period space, with the slope $R_{\rm gap} \propto P^{-0.096}$ for top-heavy and $R_{\rm gap} \propto P^{-0.089}$ for bottom-heavy core mass distributions, in good agreement with that observed \citep{vanEylen18,Martinez19}. We are also able to reproduce the rise in the gap radius with stellar mass $R_{\rm gap} \propto M_\star^{0.15}$ for top-heavy and $R_{\rm gap} \propto M_\star^{0.22}$ for bottom-heavy core mass distributions, both of which agree with the data \citep{Berger20,Petigura22} within 1-$\sigma$.

While the origin of the radius gap can be traced to primordial gas accretion, photoevaporative mass loss can further sculpt the exoplanetary population, where we see the radius gap being widened at the larger end and being filled in at the smaller end. The effect of mass loss can be seen in the radius-mass plane as well. The primordial gas accretion draws two distinct tracks in this phase space with the lower track describing bare cores and the upper track describing gas-enveloped population. Under the gas accretion model, high mass planets that assemble early reproduces Saturn-sized objects while those that assemble late reproduces the sub-Neptunes. Photoevaporation transforms some of these into massive, rocky objects that can be probed with current mass measurements. Presently, we have far less mass measurements compared to radii measurements and so it is likely that the actual mass distribution of exoplanets that make up the observed radius distribution is more bottom-heavy. In fact, the continued rise in the radius distribution below $\sim$1$R_\oplus$ shown by \citet{Hsu19} seems to suggest that this may very well be the case.

We close by listing potential observational tests. First, if the radius gap is carved out from the beginning, then we expect to find a population of rocky planets around young stars ($\lesssim$100 Myrs). Furthermore, if the radius gap can be observed at these young ages, we also expect the gap to be wider as the gas-enveloped planets would appear larger as they are still undergoing thermal cooling. Building a larger sample of planets around stars of wider age range using TESS could help test this theory, although such a test could be restricted by the current detection limit afforded by TESS around young stars \citep[e.g.,][]{Newton19,Zhou21}.

Second, we expect a population of sub-Earth size objects beyond $\sim$30--100 days. If mass loss is the original (and only) mechanism to carve out the radius valley, then we should expect significantly less planets smaller than super-Earths beyond $\sim$30 days (see, e.g., Figure 5 of \citet{Owen17} or Figures 7 and 15 of \citet{Rogers21_pevap}). Accounting for primordial gas accretion before mass loss extends the maximum period out to which we may observe the small rocky planets. \citet{Hsu19} suggest that these small planets continue to dominate the total population beyond $\sim$30 days while \citet{Wilson21} report a very slight decrease. These planets are exactly at the edge of the detection limit afforded by {\it Kepler} dataset so for more definitive measurements, we would have to wait for PLATO. 

Finally, Roman Space Telescope promises to probe the bottom of the planetary mass function at a few 100 days, with a caveat that the microlensing host stars are most likely M dwarfs. A broad mass distribution that extends to sub-Earth masses would give more credence to our model, while keeping in mind that a more peaked distribution does not necessarily rule out primordial radius gap so long as it samples core masses below those whose gas masses are expected to be limited by the maximal isothermal limit.

\facility {Exoplanet Archive}

\begin{acknowledgements}
We thank the anonymous referee whose report helped improve the manuscript. We also thank Eugene Chiang for helpful discussion on disk structures. EJL gratefully acknowledges support from NSERC, FRQNT, the McGill Space Institute, and the William Dawson Scholarship from McGill. AK was supported by the Faculty of Science Undergraduate Research Award from McGill and by the Technologies for Exo-Planetary Science CREATE program. DPT acknowledges support by the Trottier Fellowship from the Exoplanet Research Institute (iREx). This research has made use of the NASA Exoplanet Archive, which is operated by the California Institute of Technology, under contract with the National Aeronautics and Space Administration under the Exoplanet Exploration Program.
\end{acknowledgements}

\bibliography{primgap}{}
\bibliographystyle{aasjournal}

\end{document}